\documentclass[a4paper,fleqn,usenatbib]{mnras}

\usepackage{natbib}
\usepackage{soul}

\usepackage{newtxtext,newtxmath}

\usepackage[T1]{fontenc}
\usepackage{ae,aecompl}

\usepackage{todonotes}
\usepackage{outlines}
\usepackage{color}

\usepackage{graphicx}	
\usepackage{amsmath}	
\usepackage{bm}	        

\usepackage{natbib}
\newcommand{\lcdm}{\Lambda\mathrm{CDM}}
\newcommand{\avg}[1]{\left\langle #1 \right\rangle}
\def\xigg{\xi_{gg}}
\def\xigm{\xi_{gm}}

\DeclareMathOperator\erf{erf}

\newcommand{\dd}{\mathrm{d}}
\newcommand{\hmsol}{h^{-1}M_{\odot}}
\newcommand{\ms}{M_*}
\newcommand{\mh}{M_h}
\newcommand{\mpc}{\mathrm{Mpc}}
\newcommand{\hmpc}{h^{-1}\mathrm{Mpc}}

\newcommand{\hkpc}{h^{-1}\mathrm{kpc}}

\newcommand{\hhmsol}{h^{-2}M_{\odot}}
\newcommand{\ds}{\Delta\Sigma}

\newcommand{\fdet}{f_{\mathrm{det}}}

\newcommand{\sig}{\sigma_{\ln\!M_*}}

\newcommand{\kms}{\mathrm{km}\,s^{-1}}

\newcommand{\lowz}{$\mathrm{LOWZ}\,0.2{-}0.3$}
\newcommand{\cmassa}{$\mathrm{CMASS}\,0.43{-}0.51$}
\newcommand{\cmassb}{$\mathrm{CMASS}\,0.51{-}0.57$}
\newcommand{\cmassc}{$\mathrm{CMASS}\,0.57{-}0.70$}

\newcommand\redmapper{redMaPPer}

\newcommand\ihod{\texttt{iHOD}}

\newcommand{\rom}[1]{\uppercase\expandafter{\romannumeral #1\relax}}

\title[HOD of BOSS Galaxies]{On the ``Lensing is Low'' of BOSS Galaxies}
\author[Zu 2020]{
Ying  Zu$^{1}$\thanks{E-mail: yingzu@sjtu.edu.cn}
\\
$^{1}$Department of Astronomy, School of Physics and Astronomy, Shanghai Jiao Tong
University, Shanghai 200240, China\\
$^{2}$Shanghai Key Laboratory for Particle Physics and Cosmology, Shanghai Jiao Tong University, Shanghai 200240, China
}

\date{Accepted XXX. Received YYY; in original form ZZZ}

\pubyear{2020}

\begin{document}

\label{firstpage}
\pagerange{\pageref{firstpage}--\pageref{lastpage}}
\maketitle

\begin{abstract}
Recently, Leauthaud et al discovered that the small-scale lensing signal of
Baryon Oscillation Spectroscopic Survey~(BOSS) galaxies is up to 40\%
lower than predicted by the standard models of the galaxy-halo connections that
reproduced the observed galaxy stellar mass function~(SMF) and clustering. We
revisit such ``lensing is low'' discrepancy by performing a comprehensive Halo
Occupation Distribution~(HOD) modelling of the SMF, clustering, and lensing of
BOSS LOWZ and CMASS samples at Planck cosmology. We allow the
selection function of satellite galaxies to vary as a function of stellar mass
as well as halo mass. For centrals we assume their selection to depend only on
stellar mass, as informed by the directly measured detection fraction of the
redMaPPer central galaxies. The best-fitting HOD successfully describes all
three observables without over-predicting the small-scale lensing signal. This
indicates that the model places BOSS galaxies into dark matter halos of the
correct halo masses, thereby eliminating the discrepancy in the one-halo regime
where the signal-to-noise of lensing is the highest. Despite the large
uncertainties, the observed lensing amplitude above $1\,\hmpc$ remains
inconsistent with the prediction, which is however firmly anchored by the
large-scale galaxy bias measured by clustering at Planck cosmology. Therefore,
we demonstrate that the ``lensing is low'' discrepancy on scales below
$1\,\hmpc$ can be fully resolved by accounting for the halo mass dependence of
the selection function.  Lensing measurements with improved accuracy is required
on large scales to distinguish between deviations from Planck and non-linear
effects from galaxy-halo connections.
\end{abstract}




\vspace{1in}
\section{Introduction}
\label{sec:intro}

The large-scale clustering of BOSS CMASS and LOWZ galaxies are the
state-of-the-art data sets~\citep{Alam2017} that allow accurate measurements of
the expansion history of the accelerating Universe and the cosmic growth of
large-scale structures~\citep{Weinberg2013}. Such cosmological analysis with
BOSS galaxy samples relies on the accurate modelling of the galaxy-halo
connection in the Universe~\citep{wechsler2018}. However, \citet{Leauthaud2017}
recently discovered that the galaxy-galaxy lensing~(hereafter g-g lensing)
signal of the BOSS CMASS galaxies is significantly lower than predicted by their
best-fitting model that reproduced the stellar mass function~(SMF) and
clustering of those galaxies. Subsequently, \citet{Lange2019} showed that a
similar discrepancy between clustering and g-g lensing also exists in the BOSS
LOWZ galaxy sample. \citet{Leauthaud2017} explored the impact of baryonic
physics, massive neutrinos, and modifications to General Relativity~(GR), but
none of these effects can resolve this so-called ``lensing is low'' discrepancy.
Therefore, ``lensing is low' signals an alarming gap between the observed
galaxies and the dark matter haloes evolved under GR in the $\Lambda$-dominated
cold dark matter~($\Lambda$CDM) Universe described by~\citet{Planck2018}.

Beyond Planck, \citet{Leauthaud2017} and \citet{Lange2019} both found that
lowering the cosmological parameter $S_8{\equiv}\sigma_8\sqrt{\Omega_m/0.3}$ by
$2{-}3\sigma$ from the Planck value can somewhat reconcile the discrepancy.
However, as pointed out in \citet{Leauthaud2017}, the g-g lensing measurements
are dominated by non-linear scales where details of galaxy-halo connection
matter the most, a significantly lower value of $S_8$ is therefore unlikely the
favored solution unless we have thoroughly understood the systematic
uncertainties in the galaxy-halo connection at the Planck cosmology. In this
paper, we perform a comprehensive HOD modelling of the BOSS galaxies and
carefully account for the halo mass-dependence of the BOSS target selection due
to the complex colour cuts, in hopes of identifying the missing link {\it
within} the Planck $\Lambda$CDM paradigm before exploring any new physics.

Galaxy assembly bias has been regarded as one of possible solutions. In
particular, at fixed cosmology the projected clustering tightly constrains the
large-scale galaxy bias, but on small scales is severely limited by the fact
that no two fibres can be placed closer than $62^{''}$ on a given plate~(a.k.a.,
fibre collision).  Meanwhile, the g-g lensing signal is limited to scales below
$10\,\hmpc$ and primarily measures the one-halo term, which depends on the mean
halo mass of the sample and the fraction of satellite galaxies. Therefore, the
low amplitude of g-g lensing could be caused by a preference of the BOSS
galaxies to reside in halos with, e.g, lower concentration, hence the higher
large-scale bias~\citep[i.e., ``halo assembly bias'';][]{Sheth2004, Gao2005,
Gao2007, Jing2007}. However, applying two extended HOD models that include such
galaxy assembly bias effect to the CMASS sample, \citet{Yuan2020} found that
their fit still strongly suggests a ${\sim}34$ per cent discrepancy between the
projected clustering and g-g lensing signals. This finding is consistent with
the recent work of Salcedo et al. (2020), who found no evidence for a strong
galaxy assembly bias after a thorough investigation of the one and two-point
galaxy statistics in SDSS.

Another possibility is the lack of proper modelling of the spectroscopic
selection function of BOSS galaxies. Optimized for the Baryon Acoustic
Oscillation~(BAO) analysis, the BOSS target selection relies on complex sets of
colour and magnitude cuts over SDSS $ugriz$ photometry to efficiently select the
most massive and passive galaxies, at $z{<}0.43$ for LOWZ and between
$z{\sim}0.43$ and $0.7$ for CMASS, respectively~\citep{Reid2016}. To construct a
roughly ``constant mass''~(hence the name CMASS) sample, the selection criteria
are theoretically motivated by the \citet{Maraston2009} Luminous Red
Galaxy~(LRG) template that describes the passive evolution of a predominantly
metal-rich population~(with 3 per cent of the stellar mass in old metal-poor
stars). This template provides a good overall fit to the colours of massive LRGs
that generally show no evidence of additional evolution beyond
passive~\citep{Wake2006}.

If the BOSS colour and magnitude cuts closely follow the tracks of {\it all}
massive galaxies on the colour-magnitude diagram, one can then safely assume
that the detection fraction of galaxies in BOSS depends only on their stellar
mass. For instance, the best-fitting model of galaxy-halo connection adopted by
\citet{Leauthaud2017} was originally derived in \citet{Saito2016}, who performed
a joint analysis of the projected correlation function and the galaxy SMF using
subhalo abundance matching~\citep[SHAM;][]{Conroy2006, Vale2006, Shankar2006,
Guo2016}.  They account for the stellar mass incompleteness of CMASS by
down-sampling mock galaxies to match the redshift-dependent CMASS SMFs\citep[see
also][]{Rodriguez-Torres2016}.  Alternatively, \citet{Guo2018} modelled the
central and satellite galaxy completeness separately as two functions of stellar
mass, each with a three-parameter functional form proposed
in~\citet{Leauthaud2016}. Adopting the incompleteness conditional stellar mass
function of \citet{Guo2018}, \citet{Lange2019} analysed the LOWZ sample using an
analytic HOD framework, and confirmed that the LOWZ galaxies also exhibit a
discrepancy very similar to that in CMASS.

However, a simple passive evolution model is insufficient for describing at
least some of the observed massive galaxies, depending on their star formation
rates~(SFRs)~\citep{Eisenhardt1987, Runge2018, Cerulo2019}. Using the Stripe
82-Massive Galaxy Catalog~\citep[S82-MGC;][]{Bundy2015}, \citet{Bundy2017}
discovered that at the most massive end the SMF was dominated by galaxies with
some residual star formation ~($-2.7{<}\lg \mathrm{SFR}{<}-0.5$) at
$z{\sim}0.6$, which then became completely quiescent at lower redshifts. More
interestingly, there exists a sub-population of massive galaxies with ongoing
star formation($\lg \mathrm{SFR}{>}-0.5$) whose number density stays roughly
constant with redshift. Matching the BOSS catalogue to S82-MGC,
\citet{Leauthaud2016} found that many galaxies with signs of recent star
formation are excluded from the CMASS sample at redshifts between $0.4$ and
$0.6$ by the colour cuts. At $z{<}0.4$, a sliding colour cut preferentially
removed some of the relatively low-mass galaxies with bluer colours from the
LOWZ sample.

Therefore, the detection fraction of BOSS galaxies at fixed stellar mass should
also depend on halo properties that are strongly tied to galaxy colours. In
essence, this dependence is fundamentally linked to the astrophysics that
governs galaxy quenching, i.e., the rapid cessation of star
formation~\citep{Naab2017} that leads to the blue-to-red colour transformation
of galaxies. Using the colour dependence of the projected clustering and g-g
lensing of SDSS galaxies as constraint, \citet{Zu2016} explored three different
phenomenological models for galaxy quenching, which tie galaxy colours to halo
mass, stellar mass, and halo formation time, respectively. They found that the
halo mass quenching model provides an excellent fit to the SDSS measurements,
while the other two exhibit strong discrepancy between the g-g lensing and
clustering of SDSS galaxies at the high-mass end~\citep[see
also][]{Mandelbaum2016, Zu2016}, in a very similar fashion to the current
tension found within BOSS~(see their figure 10). In this paper, we build on the
\ihod{} halo quenching framework of \citet{Zu2015, Zu2016, Zu2018} and
incorporate the stellar as well as halo mass dependence of galaxy selection
functions into our analytic HOD framework, in hopes of reproducing the observed
low small-scale g-g lensing signal at fixed large-scale galaxy bias.

This paper is organized as follows. We briefly describe the BOSS LOWZ and CMASS
data and the overlapping cluster sample in \S~\ref{sec:data}, and introduce our
extended HOD model in \S~\ref{sec:method}. We present our main findings on the
lensing-clustering discrepancy in LOWZ in  \S~\ref{sec:lowz} and in CMASS in
\S~\ref{sec:cmass}. We conclude by summarizing our results and looking to the
future in \S~\ref{sec:conc}. Throughout this paper, we assume the Planck 2020
cosmology~\citet{Planck2018} with $\Omega_m{=}0.315$, $\sigma_8{=}0.811$, and
$h{=}0.6736$.  All the length and mass units in this paper are scaled as if the
Hubble constant were $100\,\kms\mpc^{-1}$. In particular, all the separations
are co-moving distances in units of $\hmpc$, and the stellar and halo mass are
in units of $\hhmsol$ and $\hmsol$, respectively. We use $\lg x{=}\log_{10} x$
for the base-$10$ logarithm and $\ln x{=}\log_{e} x$ for the natural logarithm.

\section{Data}
\label{sec:data}

\subsection{BOSS LOWZ and CMASS Samples}
\label{subsec:boss}

\begin{figure}
\begin{center}
    \includegraphics[width=0.48\textwidth]{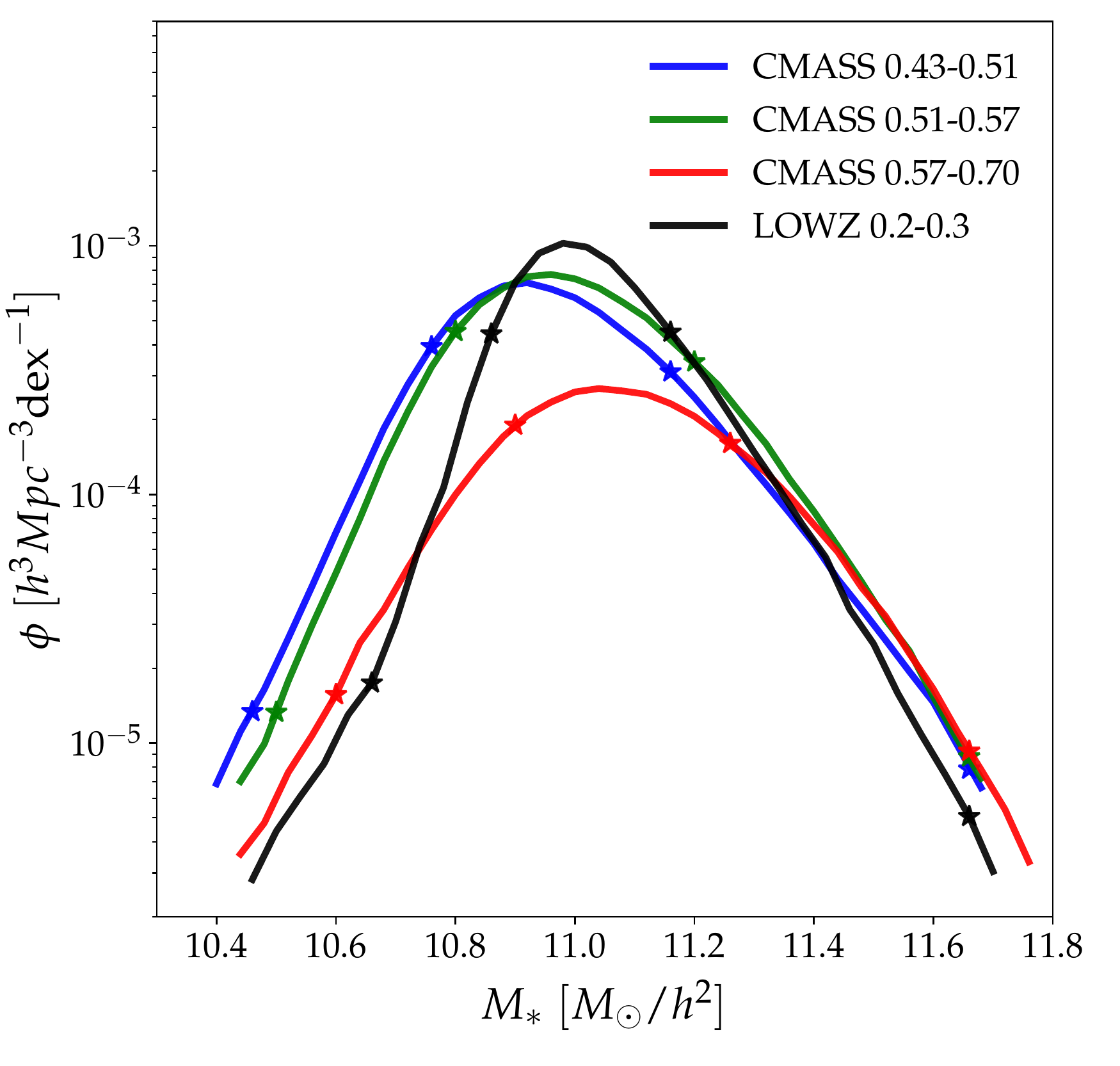} \caption{The
    stellar mass functions of the four redshift samples in our analysis,
    labelled by the legend in the top right corner. The four stars on each curve
    indicate the bin edges of the three stellar mass subsamples for which we
    measure the projected correlation functions. We carefully pick the stellar
    mass bins so that the peak of each SMF is well-contained by the intermediate
    mass bin.}
\label{fig:smf}
\end{center}
\end{figure}

As part of the SDSS-III programme~\citep{Eisenstein2011},
BOSS~\citep{Dawson2013} measured the spectra of 1.5 million galaxies over a sky
area of ${\sim}$10,000 deg$^{2}$ using the BOSS spectrographs~\citep{Smee2013,
Bolton2012} onboard the 2.5-m Sloan Foundation Telescope at the Apache Point
Observatory~\citep{Gunn1998, Gunn2006}. BOSS galaxies were selected from the
Data Release 8~\citep[DR8;][]{Aihara2011} of SDSS five-band
imaging~\citep{Fukugita1996} using two separate sets of colour and magnitude
cuts for the LOWZ~($0.15{<}z{<}0.43$) and CMASS~($0.43{<}z{<}0.7$) samples,
respectively~\citep{Reid2016}. We use the Data Release 12 of the BOSS galaxy
sample, which is also the final data release that includes the complete dataset
of the BOSS survey~\citep{Alam2015}.

Conceptually, the main difference between the LOWZ and CMASS cuts is the
extension of the CMASS selection towards the blue. Using high-resolution {\it
HST} imaging, \citet{Masters2011} showed that ${\sim}$26 per cent of the CMASS
galaxies in the COSMOS sample have late-type morphologies. Using an unbinned
maximum likelihood approach, \citet{Montero-Dorta2016} estimated that the
fraction of intrinsically blue galaxies in CMASS increases considerably as a
function of redshift, from ${\sim}36$ per cent at $z{=}0.5$ to ${\sim}46$ per
cent at $z{=}0.7$. Therefore, we anticipate a different halo mass dependence of
LOWZ and CMASS selection functions, as well as a redshift-dependence of the
parameters that describe CMASS selection.

We employ the stellar mass measurements from \citet{Chen2012}, who fit the
galaxy spectra over the rest-frame wavelength range of $3700{-}5500$ \AA{} using
a principal component analysis method. In particular, we adopt the stellar mass
estimates obtained by applying the Stellar Population Synthesis~(SPS) model of
\citet{Maraston2011} with the \citet{Kroupa2001} Initial Mass Function~(IMF) and the
dust attenuation model of \citet{Charlot2000}. All the masses are
aperture-corrected by applying the mass-to-light ratio within the fibre to the
whole galaxy. To facilitate our comparison with the results from previous
studies~\citep[e.g., ][]{Lange2019}, we follow \citet{Guo2018} by reducing the
\citeauthor{Chen2012} stellar masses by $0.155$ dex.

For the LOWZ galaxies, we will focus on the redshift range of $z{=}[0.2, 0.3]$,
for which we have a well-defined cluster catalogue from
redMaPPer~\citep{Rykoff2014}. As will be further discussed in
\S~\ref{subsec:bcg}, we rely on the volume-limited sample of brightest central
galaxies~(BCGs) defined by redMaPPer to inform us the detection fraction of
central galaxies. For the CMASS galaxies, following \citet{Leauthaud2017} we
divide them into three redshift bins, $z{=}[0.43, 0.51]$, $z{=}[0.51, 0.57]$,
and $z{=}[0.57, 0.70]$, respectively. In total, we will analyse four redshift
bins of BOSS galaxies that we will refer to as LOWZ $0.2{-}0.3$, CMASS
$0.43{-}0.51$, CMASS $0.51{-}0.57$, CMASS $0.57{-}0.70$, respectively.

Figure~\ref{fig:smf} shows the observed SMFs of the four redshift bins, marked
by the legend on the top right. The three CMASS samples all have bell-shaped
SMFs, with \cmassa and \cmassb having higher peak amplitudes than \cmassc. The
bell shape is caused by the fact that the low mass portion of the SMFs suffers
strong incompleteness, while the high mass portion enjoys relatively high
completeness. Therefore, if the selection function is additionally modulated by
halo mass, we would expect that the low stellar mass galaxies exhibit deviations
from the naive prediction assuming they are randomly selected from the parent
sample at fixed stellar mass. To capture such potential deviations in the
clustering measurements, for each redshift bin we further divide the galaxies
into three bins of stellar mass, with two bins representing the low and high
mass portions of the SMF, and the intermediate stellar mass bin anchoring the
peak of the SMF, respectively. We will refer to them simply as Low${-}M_*$,
High${-}M_*$, and Mid${-}M_*$, respectively. The bin edges we adopt are
illustrated by the star symbols on each SMF curve in Figure~\ref{fig:smf} and
listed below
\begin{itemize}
    \item \lowz{}:    $\lg\ms{=}[10.66, 10.86, 11.16, 11.66]$ ,
    \item \cmassa{}:  $\lg\ms{=}[10.46, 10.76, 11.16, 11.66]$ ,
    \item \cmassb{}:  $\lg\ms{=}[10.50, 10.80, 11.20, 11.66]$,
    \item \cmassc{}:  $\lg\ms{=}[10.60, 10.90, 11.26, 11.66]$.
\end{itemize}

Intriguingly, the \lowz{} SMF shows an additional feature compared to the CMASS
ones --- a change of slope at $\lg M_*{\sim}10.7$ towards the low mass tail.
This slope change suggests that the LOWZ SMF may consist of two distinctive
components at the low vs. high stellar mass ends, e.g., due to different
selection functions for the central and satellite galaxies. Therefore, any
successful model should explain this feature in the LOWZ SMF in addition to the
overall amplitude of the SMF.

\begin{figure*}
\begin{center}
    \includegraphics[width=0.96\textwidth]{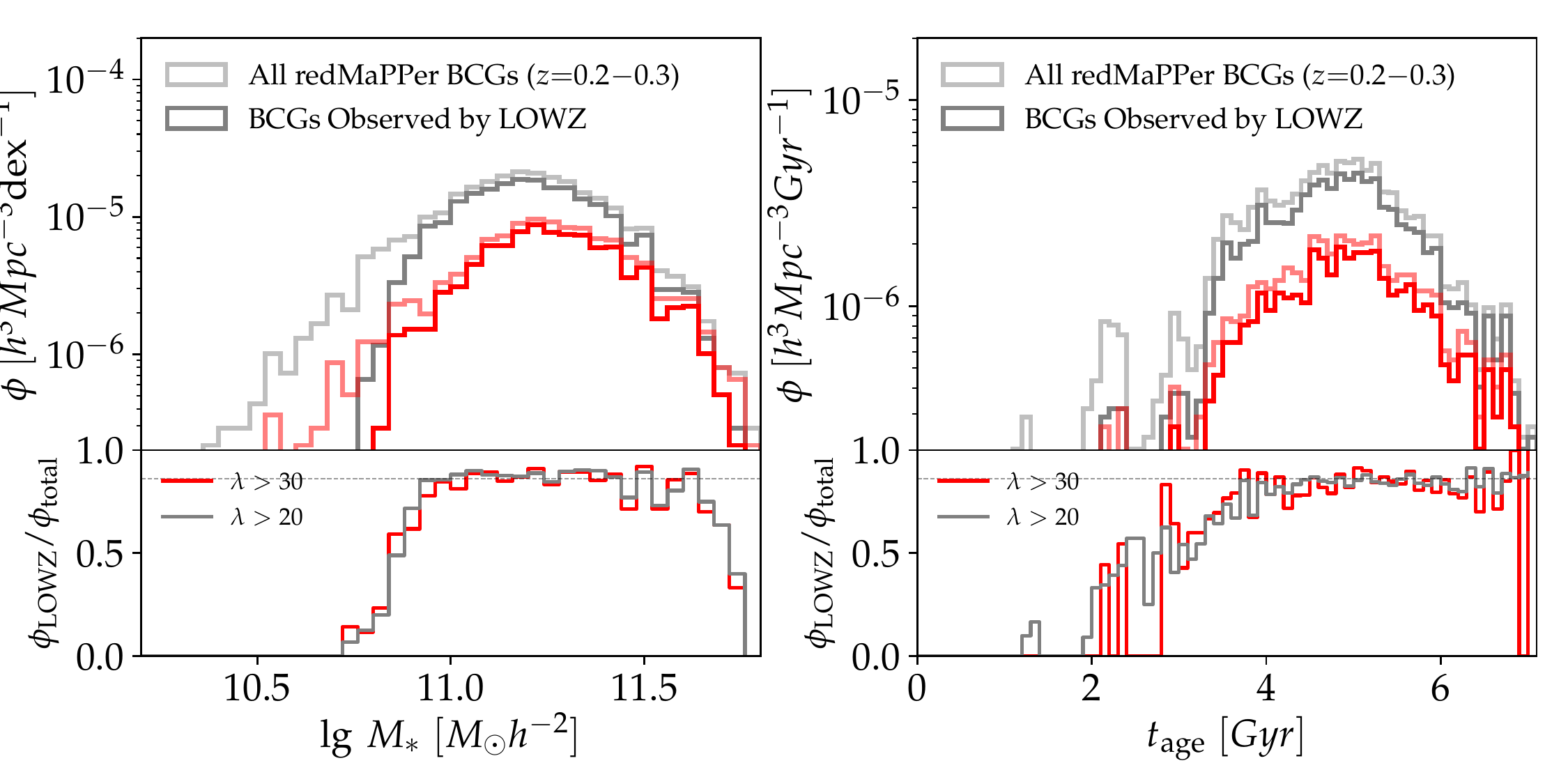}
    \caption{{\it Top}: Comparison between the number density distribution of
    BCGs of all the redMaPPer clusters~(light gray and red) and those
    spectroscopically observed by LOWZ~(dark gray and red), as functions of BCG
    stellar mass~(left) and BCG stellar age~(right), respectively. {\it Bottom}:
    The detection fraction of redMaPPer BCGs by the LOWZ spectroscopic
    observation, with the horizontal dotted line indicating the maximum LOWZ
    detection fraction of the BCGs. In each panel, gray and red histograms
    indicate the measurements from $\lambda{>}20$ and $\lambda{>}30$ clusters,
    respectively.  The selection functions of BCGs show little dependence on
    cluster richness.}
\label{fig:clslowz}
\end{center}
\end{figure*}

We emphasize that the deliberate binning of stellar mass is crucial to our HOD
analysis. In particular, the large-scale clustering of galaxies measures the
weighted average of the central and satellite galaxy biases. By dividing the
galaxies into stellar mass bins with drastically different levels of
completeness, the large-scale clustering thus provides key information on the
satellite fraction as a function of both stellar mass and completeness, helping
disentangle the small-scale g-g lensing signal into central vs. satellite
contributions. In comparison, \citet{Leauthaud2017} used the projected
clustering of a single bin of all CMASS galaxies, while \citet{Lange2019}
adopted two somewhat arbitrary stellar mass bins regardless of where the
observed SMF peaks.

\subsection{Brightest Central Galaxy Sample from redMaPPer Cluster Catalogue}
\label{subsec:bcg}

To investigate the BOSS detection fraction of the central galaxies, we employ an
SDSS cluster catalogue derived from SDSS DR8 imaging using the
red-sequence-based matched-filter photometric cluster finding algorithm
redMaPPer~\citep{Rykoff2014}. The redMaPPer cluster catalogue measures the
richness $\lambda$ of satellite galaxies brighter than $0.2\,L_*$ within an
aperture ${\sim}1\,\hmpc$~(with a weak dependence on $\lambda$) as its proxy for
halo mass. \citet{Rykoff2014} demonstrated that the cluster catalogue is
approximately volume-complete above $\lambda{=}20$ to $z{\simeq}0.33$ with
excellent photometric redshift estimates~($\delta(z)=0.006/(1+z)$). Therefore,
the redMaPPer BCG sample serve as an ideal data set that allows us to {\it
directly} measure the BOSS detection fraction of the central galaxies of massive
haloes as a function of both stellar mass and cluster richness.

We employ $3822$ BCGs of the redMaPPer $\lambda{>}20$ clusters between
$z{=}0.2{-}0.3$ and within the same sky area covered by the \lowz{} sample. The
average halo mass of those clusters is accurately measured from weak
lensing~\citep[${\simeq}1.86\times 10^{14}\hmsol$;][]{Miyatake2016, Simet2017},
and their large-scale halo bias is consistent with the lensing mass.  When
divided by galaxy concentration, the halo assembly bias exhibited by those
clusters is also in perfect agreement with that predicted by the $\Lambda$CDM at
Planck cosmology~\citep{Zu2017}. Therefore, at the high stellar mass end, there
is no evidence an analogous ``lensing is low'' discrepancy~(i.e., halo mass-bias
discrepancy) within the volume-limited sample of redMaPPer BCGs.

However, due to the lack of spectra for the BCGs that were not observed by BOSS~($788$; 21\% of the
total sample), we do not have \citet{Chen2012} stellar mass estimates available
for these objects. Thanks to the excellent photometric redshift estimates by the
redMaPPer algorithm, we can derive reasonably accurate stellar masses for all
the BCGs by fitting a two-component Simple Stellar Population~(SSP) template to
their SDSS {\it ugriz} photometry. Following \citet{Maraston2009}, we include a
dominant component~(97 per cent) of a solar metallicity population, with 3 per
cent of a metal-poor~($Z{=}0.008$) population of the same age. We utilized the
\texttt{EzGal} software~\citep{Mancone2012} and adopt the \citet{Bruzual2003}
SSP model and a \citet{Chabrier2003} IMF for the fits.  For the aperture
correction, we carry out the fit on extinction-corrected model magnitudes that
are scaled to the $i$-band $c$-model magnitudes.

The SSP and IMF assumptions are different from the \citet{Chen2012} fits, so the
BCG stellar masses inferred from \texttt{EzGal} are systematically lower than
the original \citet{Chen2012} values by ${\sim}0.2$ dex, hence ${\sim}0.045$ dex
lower than the values adopted in our analysis~(with the $0.155$-dex offset).
There is also a scatter of ${\sim}0.1$ dex between the \texttt{EzGal} and
\citet{Chen2012} masses, mainly due to the difference in star formation
histories. The difference between the two stellar mass estimates are not
important because our goal in this Section is to investigate the functional
forms of the BOSS selection function and its dependence on the host halo mass
proxy $\lambda$, which should be independent on the exact values of the stellar
mass estimates.

The top panels of Figure~\ref{fig:clslowz} show the cluster number density
distributions as functions of stellar mass~(left) and stellar age~(right) of the
BCGs. In each panel, the gray histograms show the results for all the clusters
with $\lambda$ above $20$, while the red histograms with $\lambda$ above $30$.
Within each colour, the light and dark-colour histograms indicate the
distributions of all BCGs and those spectroscopically observed by LOWZ,
respectively. The bottom panels show the LOWZ detection fraction of all the BCGs
as the ratios of the dark over light-colour histograms.

For the detection fraction as a function of stellar
mass on the bottom left panel of Figure~\ref{fig:clslowz}, there is a plateau of maximum detection efficiency of ${\sim}86\%$ at
$11{<}\lg M_*{<}11.4$~(\texttt{EzGal} mass estimates), which declines towards
zero at both ends of the stellar mass distribution. The plateau is probably due
to some photometric effects like the masking that reduces the effective area of
the LOWZ target selection. The cut-offs can be both described by the functional
forms introduced by \citet{Leauthaud2016}, as shown by the gray dashed curve in
the bottom panel of Figure~\ref{fig:fdet}, with some slight shift due to the
difference between the \citet{Chen2012} and \texttt{EzGal} stellar mass
estimates. The low-mass cut-off is relatively sharp, indicating that the LOWZ
target selection does a great job selecting a stellar mass-thresholded sample of
central galaxies. More important, the cut-off is roughly independent of
$\lambda$~(compare red to gray), indicating that the selection is insensitive to
halo mass, at least in the cluster mass regime.  Meanwhile, it is unclear what
caused the cut-off at the high-mass end, probably due to some rejuvenated star
formation in the most massive systems~\citep{Runge2018}. The impact on our
analysis is however negligible because the number of galaxies affected by this
high-mass cut-off is very low.

The bottom right panel of Figure~\ref{fig:clslowz} shows the two detection
fractions as functions of stellar age for the $\lambda{>}20$~(gray) and
$\lambda{>}30$~(red) samples. The stellar age of each BCG is derived during the
\texttt{EzGal} fitting to broad-band colours assuming an SSP at some birth
redshift, and is therefore more meaningful when used as a measure of the
relative age among the stellar population than absolute. Similar to the stellar
mass dependence, the stellar age dependence also reaches a maximum detection
efficiency of ${\sim}86\%$ at $t_{\mathrm{age}}{>}5\,\mathrm{G}yrs$, and slowly declines into zero
across of spread of $\Delta\,t_{\mathrm{age}}{\sim}4\,\mathrm{G}yrs$. This
confirms our expectation from \S~\ref{sec:intro} that the incompleteness is
primarily due to the blueward deviation from the colour of a passively evolving
galaxy due to star formation. Such deviation is also insensitive to halo mass,
as the rich clusters exhibit similar stellar-age selection effects than the poor
ones~(compare red to gray).

To summarize, the selection function of central galaxies of massive
clusters reaches a maximum of ${\sim}86\%$ for the high-stellar mass, old-age population,
with a sharp cut-off at $\lg M_*{\sim}10.9$ that unfolds into a slow decline
across a span in stellar age of $4\,\mathrm{G}yrs$. More important, such
behavior is relatively insensitive to cluster richness, hence unlikely a strong
function of halo mass. Therefore, in our analysis with both the LOWZ and CMASS
samples, we will extrapolate our findings among the redMaPPer clusters into the
low mass regime and assume that the detection fraction of central galaxies is a
function of only stellar mass.

\subsection{Large-scale Structure Measurements}
\label{subsec:measurement}

For each of the redshift bins, we jointly analyse the observed
SMF~(Figure~\ref{fig:smf}), the g-g lensing of the total galaxy sample, and the
projected correlation functions of the three stellar mass subsamples defined in
\S~\ref{subsec:boss}~(also see the stars in Figure~\ref{fig:smf}). We employ the
g-g lensing signal of the \lowz{} sample measured by \citet{Lange2019}, and that
of the \cmassa{}, \cmassb{}, and \cmassc{} samples measured by
\citet{Leauthaud2017}, respectively. Below we briefly describe the measurements
of the projected auto-correlation functions for the stellar mass subsamples, and
refer readers to \citet{Leauthaud2017} and \citet{Lange2019} for the technical
details of the g-g lensing measurements.

We measure the projected auto-correlation function $w_p$ as the integration of
the redshift-space correlation function $\xi^{rs}(r_p, r_\pi)$ as follows
\begin{equation}
    w_p = \int_{-r_{\pi}^{\mathrm{max}}}^{+r_{\pi}^{\mathrm{max}}}\!
    \xi^{rs}(r_p, r_\pi)\,\mathrm{d}r_{\pi},
    \label{eqn:wp}
\end{equation}
where $r_p$ and $r_{\pi}$ are the projected and line-of-sight separation of
galaxy pairs, respectively, and $r_{\pi}^{\mathrm{max}}$ is the integration
limit along the line of sight, which we set to be $100\,\hmpc$ to minimize the
redshift space distortion effects. To make sure that the spatial correlation
signal is from a contiguous region on the sky, we only use the BOSS observations
of the North Galactic Cap. We adopt the Davis-Peebles
estimator~\citep{Davis1983} for our correlation measurements, so that
\begin{equation}
    \xi^{rs}(r_p, r_\pi) = \frac{DD}{DR} - 1,
\end{equation}
where $DD$ and $DR$ represent the number counts of pairs of two data galaxies,
one data and one random galaxies, respectively.  We have also computed $w_p$
with the Landy-Szalay estimator~\citep{Landy1993} and the results are similar on
all scales except above $30\,\hmpc$, where the Landy-Szalay measurements show a
slightly shallower slope than suggested by the matter clustering predicted by
Planck on relevant scales. We have tested our analysis using the Landy-Szalay
measurements and confirmed that the impact on our results are negligible.

As mentioned in \S~\ref{sec:intro}, the BOSS spectroscopic observation is
subject to the fibre collision effect, so that galaxy pairs close to $62^{''}$
were severely under-sampled. Such effect can be partially remedied by
sophisticated schemes that either takes advantage of the multiple passes of the
survey~\citep{Guo2012}, or the probabilistic distribution of close galaxies
along the line of sight~\citep{Yang2019}. However, those schemes were most
intensively tested against a much denser sample~(e.g., the SDSS Main Galaxy
Sample), and it is unclear how robust the correction would work for BOSS. We
therefore decide to apply the simplest nearest-neighbour correction and use
distance scales above the fibre collision scale for our correlation function
analysis. In particular, the maximum fibre-collided scale that corresponds to
the maximum redshift of each of our four redshifts bins are $0.25\,\hmpc$,
$0.40\,\hmpc$, $0.44\,\hmpc$, $0.52\,\hmpc$, respectively, and the minimum
distance of our analysis is $0.6\,\hmpc$ across all redshift bins. For the g-g
lensing signals, following \citet{Lange2019} we limit our fits to scales above
$100\,\hkpc$, as on smaller scales the signals are strongly affected by the
contributions from the galaxy stellar mass and the subhalo mass associated with
the satellite galaxies~\citep{Zu2015}.

\section{Methodology}
\label{sec:method}

We adopt an analytic HOD model to predict the number density distributions,
projected auto-correlation functions, and g-g lensing signals of BOSS galaxies.
Compared to studies using the simulation-based methods~\citep{Leauthaud2017,
Guo2018}, our method is computationally efficient when exploring the parameter
space, but relatively lacking in accuracy, especially in the one-to-two-halo
transition regime where the triaxial halo shape~\citep{Jing2002} and halo
exclusion becomes important~\citep{Tinker2005, Zu2014, Garcia2019}. Fortunately,
our analysis is insensitive to the systematic uncertainties at the transition
scales, as the statistical uncertainties of the current g-g lensing signals
at those scales are similarly large~(${\sim}20{-}30\%$), and we primarily rely on the
scales above~(clustering) and below~(lensing) for our constraints. However, we
expect that an emulator-based method similar to that of \citep{Wibking2019,
Zhai2019, Nishimichi2019, Wibking2020} is necessary for exploring the
cosmological information within the clustering and lensing of the BOSS galaxies
and in future surveys.

The HOD model in our analysis was heavily based on the \ihod{} framework
developed in a series of papers~\citep{Zu2015, Zu2016}, which drew insights from
earlier works of \citet{Berlind2002, Guzik2002, Tinker2005, Mandelbaum2006,
Zheng2007, Yoo2006, Leauthaud2011}. We briefly describe the HOD prescription,
with a focus on the modifications to the original \ihod{} framework, and refer
the interested readers to the aforementioned two \ihod{} papers for details.

\subsection{Halo Occupation Distribution}
\label{subsec:hod}

We start from the joint probability density distribution~(PDF) of galaxy stellar
mass $\ms$ and host halo mass $\mh$,
\begin{equation}
    p(\ms, \mh) = \frac{\lg e}{\ms\,n_g}\frac{\mathrm{d}N(\ms|\mh)}{\mathrm{d}\lg\ms}\frac{\mathrm{d}n}{\mathrm{d}\mh},
\label{eqn:p2d}
\end{equation}
where $\mathrm{d} N(\ms |\mh)/\mathrm{d}\lg\ms$ is the total number of
galaxies~(both observed and unobserved) per dex in stellar mass within halos at
given mass $\mh$, $n_g$ is the total galaxy number density, and
$\mathrm{d}n/\mathrm{d}\mh$ is the halo mass function at Planck cosmology.
Following \citet{Zu2015}, we hereafter refer to $\mathrm{d} N(\ms
|\mh)/\mathrm{d}\lg\ms$ as $\avg{N(\ms |\mh)}$, assuming a fixed bin in
logarithmic stellar mass.

The main advantage of starting with $p(\ms, \mh)$ is that, for an observed
population of galaxies, whether it be quenched galaxies~\citep{Zu2016} or BOSS
observed galaxies as we study here, the 2D PDF of such population can be
directly obtained by multiplying $p(\ms, \mh)$ and the 2D selection function
$\fdet(\ms, \mh)$. In the case of studying galaxy quenching in \citet{Zu2016},
the 2D selection function is simply the 2D quenched fraction of galaxies as a
function of $\ms$ and $\mh$.

Our analytic model for $\avg{N(\ms|\mh)}$ has two components: 1) the mean and
the scatter of the Stellar-to-Halo Mass Relation~(SHMR) for the central
galaxies, the combination of which automatically specifies
$\avg{N_{\mathrm{cen}}(\ms|\mh)}$, and 2) the mean number of satellite galaxies
with stellar mass $\ms$ inside halos of mass $\mh$,
$\avg{N_{\mathrm{sat}}(\ms|\mh)}$. We adopt the same parameterisation for the
two components as in \citet{Zu2015}.

At fixed halo mass, we assume a log-normal probability distribution for the
stellar mass of the central galaxies, hence a log-normal scatter. The mean SHMR
is then the sliding mean of the log-normal distribution as a function of the
halo mass, $f_{\mathrm{SHMR}}{\equiv}\exp\avg{\ln\ms(\mh)}$. We adopt a
functional form for $f_{\mathrm{SHMR}}$ proposed by \citet{Behroozi2010} via its
inverse function\footnote{There was a typo in the equation 19 of \citet{Zu2015}:
$\exp\rightarrow 10$},
\begin{equation}
    \mh = M_1  m^\beta 10^{\left(m^\delta / (1 + m^{-\gamma}) - 1/2\right)},
    \label{eqn:shmr}
\end{equation}
where $m\equiv\ms/M_{*,0}$. Among the five parameters that describe
$f_{\mathrm{SHMR}}$, $M_1$ and $M_{*,0}$ are the characteristic halo mass and
stellar mass that separate the behaviours in the low and high mass
ends~($f_{\mathrm{SHMR}}(M_1){=}\ln M_{*,0}$). The inverse function starts with
a low-mass end slope $\beta$, crosses a transitional regime around ($M_{*,0}$,
$M_1$) dictated by $\gamma$, and reaches a high-mass end slope $\beta+\delta$.
For the log-normal scatter, we keep the scatter independent of halo mass below
$M_1$, but allow more freedom in the scatter above the characteristic mass
scale, with an extra component that is linear in $\lg\mh$:
\begin{equation}
    \sig(\mh) = \left\{ \begin{array}{ll}
        \sig,&\mbox{ $\mh<M_1$} \\
            \sig + \eta \lg \frac{\mh}{M_1},&\mbox{ $\mh\geq M_1$}
\end{array} \right.
\label{eqn:sigmh}
\end{equation}
For the satellite populations, we model the expectation value of the satellite
occupation $\avg{N_{\mathrm{sat}}(\ms|\mh)}$ as the derivative of the satellite
occupation number in stellar mass-thresholded samples,
$\avg{N_{\mathrm{sat}}(>\ms|\mh)}$, which is parameterised as a power of halo
mass and scaled to $\avg{N_{cen}(>\ms|\mh)}$ as follows,
\begin{equation}
    \avg{N_{\mathrm{sat}}(>\ms|\mh)} = \avg{N_{\mathrm{cen}}(>\ms|\mh)}
    \left(\frac{\mh}{M_{\mathrm{sat}}}\right)^{\alpha_{\mathrm{sat}}}.
    \label{eqn:nsat}
\end{equation}
We do not include the exponential cut-off in the equation 22 of \citet{Zu2015},
which has negligible impact on our constraints. We parameterise the
characteristic mass of a single satellite-hosting halo $M_{\mathrm{sat}}$ as a
simple power law function of the threshold stellar mass, so that
\begin{equation}
    \frac{M_{\mathrm{sat}}}{10^{12}\hhmsol} = B_{\mathrm{sat}}
    \left(\frac{f_{\mathrm{SHMR}}^{-1}(\ms)}{10^{12}\hhmsol}\right)^{\beta_{\mathrm{sat}}}.
\end{equation}
In practice, we choose a $0.02$ dex bin size in stellar mass for the numerical
differentiation of $\avg{N_{\mathrm{sat}}(>\ms|\mh)}$.

For any given sample defined between $\ms^{\mathrm{min}}$ and
$\ms^{\mathrm{max}}$, the PDF of satellite occupation was commonly assumed to be
Poisson with a mean of
$\avg{N_{\mathrm{sat}}(\mh)}{\equiv}\avg{N_{\mathrm{sat}}(\ms^{\mathrm{min}}<\ms<\ms^{\mathrm{max}}|\mh)}$,
so that the expected total number of satellite pairs within a halo of mass $\mh$
is $\avg{N_{\mathrm{sat}}(\mh) \left(N_{\mathrm{sat}}(\mh) -
1\right)}{=}\avg{N_{\mathrm{sat}}(\mh)}^2$~\citep{Berlind2002}.  However, in the
BOSS sample we anticipate that the distribution of satellite occupation is
narrower than Poisson: the underlying population of massive LRGs is more
concentrated in the inner halo region than the average galaxies due to dynamical
friction~\citep{Chandrasekhar1943}, but the fibre collision effect would remove
many of the close pairs of LRGs to even out the observed satellite occupation
numbers across halos of the same mass.  We test this ansatz by measuring the
number of close neighbours $N_{\mathrm{close}}$ around massive LRGs that are at
least 50 per cent more massive than the second massive neighbour within a
projected distance of $1\,\hmpc$ and a line-of-sight separation of
$\Delta\,v=\pm500\,\mathrm{km}/s$~\citep[see ``locally brightest galaxies''
in][]{Anderson2015, Mandelbaum2016}. Among the systems with at least one close
neighbour, $N_{\mathrm{close}}$ is predominantly unity, with a few per cent of
them having $N_{\mathrm{close}}{\geq}2$. Therefore, we adopt a satellite
occupation distribution similar to the model ``Average'' described in
\citet{Berlind2002}, so that if $\avg{N_{\mathrm{sat}}(\mh)}$ is between
integers $i$ and $j{=}i+1$, the halo has $p{=}\avg{N_{\mathrm{sat}}(\mh)}{-}i$
chance of hosting $j$ satellites, and $1{-}p{=}j{-}\avg{N_{\mathrm{sat}}(\mh)}$
chance of hosting one fewer satellites. Therefore, the expectation number of
satellite pairs within a halo of mass $\mh$ is $2\avg{N_{\mathrm{sat}}(\mh)}
i{-}i j$, substantially smaller than the Poisson expectation of
$\avg{N_{\mathrm{sat}}(\mh)}^2$ when $\avg{N_{\mathrm{sat}}(\mh)}{\leq}2$.

In order to model the spatial distribution of galaxies within halos, we assume
the isotropic Navarro-Frenk-White~\citep[NFW:][]{Navarro1997} density profile
for halos with the concentration--mass relation $c_{\mathrm{dm}}(\mh)$
calibrated by~\citet{Zhao2009}. We place central galaxies at the barycentres of
the NFW halos, and assume an NFW profile for the satellite distribution, but
with a different amplitude of the concentration--mass relation than the dark
matter. In particular, we set $c_{\mathrm{sat}}(\mh)\equiv f_{c} \times
c_{\mathrm{dm}}(\mh)$, where $f_c$ characterises the spatial distribution of
satellite galaxies relative to the dark matter within halos.

Armed with the PDFs of the central and satellite occupations of each galaxy
sample, we can predict the 3D real--space galaxy auto--correlation function
$\xigg$ and the galaxy--matter cross--correlation function $\xigm$, using the
halo mass and bias functions predicted at the median redshift of the sample for
the Planck 2020 cosmology. The technical details of this prediction can be found
in \citet{Zu2015}. The signals of $w_p$ and $\ds$ are then obtained by
projecting $\xigg$ and  $\xigm$ along the line of sight, respectively. The
projection of $\xigg$ to $w_p$ is given by Equation~\eqref{eqn:wp}, while for
the g-g lensing it is via
\begin{equation}
    \ds(r_p) = \langle\Sigma(<r_p)\rangle - \Sigma(r_p),
\label{eqn:ds}
\end{equation}
where
\begin{equation}
   \Sigma(r_p) =  \bar{\rho}_m \int_{-\infty}^{+\infty}\! \left[1 + \xigm(r_p, r_\pi)\right]\,\dd r_\pi ,
\label{eqn:ds2}
\end{equation}
and
\begin{equation}
   \langle\Sigma(<r_p)\rangle = \frac{2}{r_p^2} \int_0^{r_p}\! r_p^\prime \Sigma(r_p^\prime)\,\dd r_p^{\prime}.
\label{eqn:ds1}
\end{equation}
When computing the predictions for $w_p$, we apply a correction for the residual
redshift-space distortion effect that causes an enhancement of $w_p$ on large
scales due to the non-zero pairwise velocity between galaxies separated beyond
$r_{\pi}^{\mathrm{max}}{=}100\,\hmpc$, following the recipe of
\citet{vandenBosch2013}. Meanwhile for predicting $\ds$, we do not include the
contributions from the galaxy stellar mass and the subhalo mass as was done in
\citet{Zu2015}, because we have limited our fit to scales above $0.1\,\hmpc$
where these two terms are sub-dominant.

Finally, the galaxy SMF can be obtained simply as
\begin{equation}
\Phi(\ms) = n_g \int_0^{+\infty} \! p(\ms, \mh)\,\fdet(\ms, \mh)\,\mathrm{d}\mh,
\label{eqn:smf}
\end{equation}
where the 2D selection function $\fdet(\ms, \mh)$ will be modelled separately
for centrals and satellites~(as will be described in \S~\ref{subsec:selection}).

To summarise the standard HOD prescription, we have 11 model parameters so far.
Among them $\{\lg\mh^1, \lg\ms^0, \beta, \delta, \gamma\}$ describe the mean
SHMR, $\{B_{\mathrm{sat}}, \beta_{\mathrm{sat}}, \alpha_{\mathrm{sat}}\}$
describe the parent HOD of satellite galaxies, $\{\sig, \eta\}$ describe the
logarithmic scatter about the mean SHMR, and $f_c$ is the ratio between the
concentrations of the satellite distribution and the dark matter profile.

\subsection{2D Selection Function of BOSS Galaxies}
\label{subsec:selection}

\begin{figure}
\begin{center}
    \includegraphics[width=0.48\textwidth]{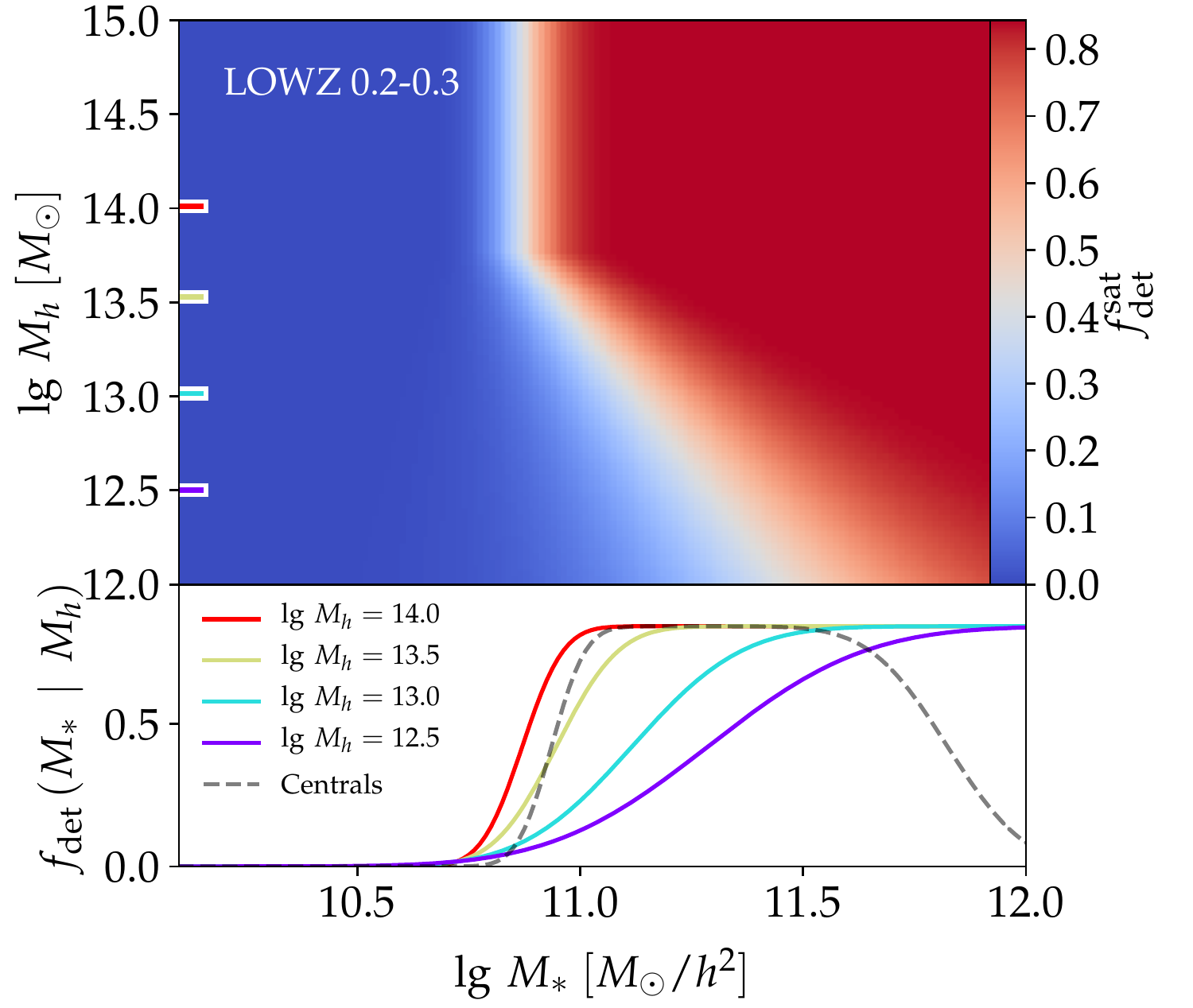} \caption{{\it Top}: The best-fitting 2D selection function of satellite galaxies in the
    \lowz{} sample on the $\lg\ms$ vs.  $\lg\mh$ plane, colour-coded by the
    colour bar on the right. {\it Bottom}: The 1D satellite selection functions
    within haloes of four different masses~(solid curves) and the central
    selection function predicted for the \lowz{} sample~(gray dashed curve) as
    described by Equation~\ref{eqn:fdet}. The \lowz{} sample preferentially
    selected satellite galaxies from the higher mass haloes below
    $\mh{=}13.76$.}
\label{fig:fdet}
\end{center}
\end{figure}

In the absence of any selection functions, the HOD prescription described in
\S~\ref{subsec:hod} would be adequate for predicting the measured SMF, $w_p$,
and $\ds$ for volume-limited stellar mass samples. As introduced in
\S~\ref{sec:intro} and ~\S~\ref{sec:data}, the BOSS colour and magnitude cuts
have introduced a complex selection function that we will model as a 2D function
of $\ms$ and $\mh$. Motivated by the findings in \citep{Zu2016}, we model the
central and satellite selection functions separately, as the quenched fractions
of centrals vs. satellites exhibit different dependences on halo mass.

\begin{table}
\renewcommand*{\arraystretch}{1.3}
\centering \caption{Posterior constraints of the model parameters for the four redshift samples. The uncertainties are the $68\%$ confidence regions derived from the 1D posterior probability distributions.}
\begin{tabular}{ccccc}
\hline
\hline
    Parameter & LOWZ & CMASS & CMASS & CMASS \\
          & $0.2{-}0.3$ &$0.43{-}0.51$ &$0.51{-}0.57$ & $0.57{-}0.70$ \\
\hline
 $\lg\mh^1$              & $13.00_{-0.32}^{+0.29}$ & $13.08_{-0.33}^{+0.27}$  &  $13.35_{-0.17}^{+0.13}$ &  $13.41_{-0.27}^{+0.22}$  \\
 $\lg\ms^0$              & $10.76_{-0.13}^{+0.12}$ & $10.71_{-0.20}^{+0.14}$  &  $10.86_{-0.07}^{+0.06}$ &  $10.99_{-0.11}^{+0.08}$  \\
 $\delta$                & $\;\;0.38_{-0.08}^{+0.05}$  & $\;\;0.85_{-0.14}^{+0.17}$   &  $\;\;1.19_{-0.18}^{+0.19}$  &  $\;\;1.10_{-0.23}^{+0.23}$   \\
 $\beta$                 & $\;\;0.74_{-0.17}^{+0.19}$  & $\;\;0.71_{-0.12}^{+0.17}$   &  $\;\;0.72_{-0.09}^{+0.17}$  &  $\;\;0.80_{-0.14}^{+0.18}$   \\
 $\gamma$                & $\;\;2.61_{-0.59}^{+0.57}$  & $\;\;1.35_{-0.56}^{+0.55}$   &  $\;\;2.92_{-0.94}^{+1.08}$  &  $\;\;1.96_{-0.84}^{+0.89}$   \\
 $\sig$                  & $\;\;0.42_{-0.03}^{+0.03}$  & $\;\;0.47_{-0.04}^{+0.04}$   &  $\;\;0.50_{-0.03}^{+0.04}$  &  $\;\;0.49_{-0.04}^{+0.04}$   \\
 $\eta$                  & $\;\;0.01_{-0.01}^{+0.01}$  & $\;\;0.39_{-0.08}^{+0.09}$   &  $\;\;0.48_{-0.12}^{+0.14}$  &  $\;\;0.26_{-0.08}^{+0.11}$   \\
 $B_{\mathrm{sat}}$      & $\;\;5.53_{-1.66}^{+2.08}$  & $\;\;5.62_{-1.92}^{+2.34}$   &  $\;\;3.60_{-1.30}^{+1.57}$  &  $\;\;9.32_{-3.28}^{+3.87}$   \\
 $\beta_{\mathrm{sat}}$  & $\;\;0.94_{-0.09}^{+0.11}$  & $\;\;0.36_{-0.10}^{+0.11}$   &  $\;\;0.24_{-0.05}^{+0.06}$  &  $\;\;0.24_{-0.08}^{+0.09}$   \\
 $f_c$                   & $\;\;0.38_{-0.18}^{+0.26}$  & $\;\;0.68_{-0.27}^{+0.32}$   &  $\;\;3.38_{-1.29}^{+1.58}$  &  $\;\;1.13_{-0.48}^{+0.55}$   \\
 $\alpha_{\mathrm{sat}}$ & $\;\;1.00_{-0.03}^{+0.03}$  & $\;\;0.99_{-0.04}^{+0.03}$   &  $\;\;0.97_{-0.03}^{+0.03}$  &  $\;\;0.99_{-0.03}^{+0.03}$   \\
 $\lg  M_{*}^c$          & $10.94_{-0.01}^{+0.01}$ & $10.96_{-0.07}^{+0.09}$  &  $11.25_{-0.15}^{+0.22}$ &  $11.44_{-0.12}^{+0.17}$  \\
 $\sigma^c$              & $\;\;0.09_{-0.01}^{+0.01}$  & $\;\;0.21_{-0.05}^{+0.05}$   &  $\;\;0.27_{-0.08}^{+0.09}$  &  $\;\;0.42_{-0.07}^{+0.07}$   \\
    $\mu$                &\hspace{-3.5pt} $-0.33_{-0.13}^{+0.11}$ & $\;\;0.24_{-0.06}^{+0.07}$   &  $\;\;0.38_{-0.07}^{+0.08}$  &  $\;\;0.36_{-0.09}^{+0.10}$   \\
    $\nu$                &\hspace{-3.5pt} $-0.23_{-0.06}^{+0.06}$ & $\;\;0.30_{-0.06}^{+0.08}$   &  $\;\;0.35_{-0.05}^{+0.06}$  &  $\;\;0.35_{-0.07}^{+0.09}$   \\
 $\lg  M_{*}^s$          & $10.88_{-0.03}^{+0.03}$ & $11.39_{-0.17}^{+0.19}$  &  $11.96_{-0.28}^{+0.28}$ &  $11.86_{-0.25}^{+0.28}$  \\
 $\lg  M_{h}^s$          & $13.76_{-0.24}^{+0.28}$ & $15.01_{-0.33}^{+0.59}$  &  $15.44_{-0.48}^{+0.76}$ &  $14.87_{-0.48}^{+0.70}$  \\
 $\sigma^s$              & $\;\;0.10_{-0.02}^{+0.02}$  & $\;\;0.76_{-0.13}^{+0.16}$   &  $\;\;1.09_{-0.19}^{+0.25}$  &  $\;\;1.00_{-0.20}^{+0.24}$   \\
 $f_{\mathrm{max}}$      & $\;\;0.84_{-0.05}^{+0.06}$  & $\;\;0.84_{-0.05}^{+0.05}$   &  $\;\;0.86_{-0.06}^{+0.06}$  &  $\;\;0.86_{-0.05}^{+0.06}$   \\
\hline
\end{tabular}
\label{tab:constraints}
\end{table}

For the central galaxies, based on the findings in~\S~\ref{subsec:bcg} we
conclude that the selection is largely independent on halo mass, so that the
detection fraction can be reduced into a single-parameter function
$\fdet^{\mathrm{cen}}(\ms)$. Informed by the measurements in
Figure~\ref{fig:clslowz} and inspired by the functional form of
\citet{Leauthaud2016}, we model the central detection fraction as
\begin{equation}
\fdet^{\mathrm{cen}}(\ms) = \frac{f_{\mathrm{max}}}{2} \left( 1 + \erf \left[ (\lg M_* -
    \lg M_{*}^{c}) / \sigma^{c}\right] \right) f_{\mathrm{cut}}(M_*),
    \label{eqn:fdet}
\end{equation}
where $f_{\mathrm{max}}$ is the maximum detection fraction, $\lg M_{*}^{c}$ is
the characteristic stellar mass at which $\fdet^{\mathrm{cen}}$ drops to
$f_{\mathrm{max}}/2$, $\sigma^{c}$ dictates the width of the decline into zero,
and $f_{\mathrm{cut}}(\lg M_*)$ describes the cut-off at the high stellar mass
end. For the \lowz{} sample, we fix the cut-off based on the
Figure~\ref{fig:clslowz} results
\begin{equation}
    f_{\mathrm{cut}}(\ms) =\frac{1}{2}\left(1 -  \erf \left[(\lg M_* - 11.83) /
    0.194\right]\right),
    \label{eqn:fcut}
\end{equation}
while for the CMASS samples we set $f_{\mathrm{cut}}(\ms)=1$.

\begin{figure*}
\begin{center}
    \includegraphics[width=0.96\textwidth]{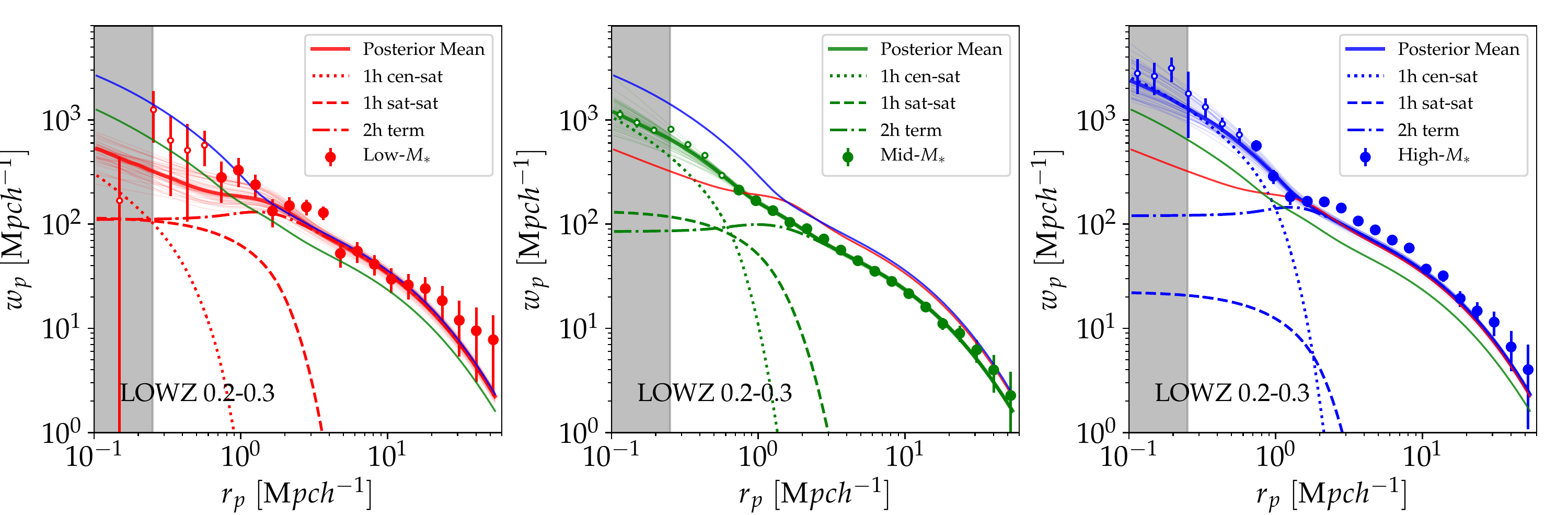}
    \caption{Comparison between the projected auto-correlation functions $w_p$
    measured from data and predicted from the posterior mean model, for the
    Low${-}M_*$~(left), Mid${-}M_*$~(middle), and High${-}M_*$~(middle) stellar
    mass subsamples of the \lowz{} galaxies. In each panel, large filled circles
    with errorbars are the measurements used by our analysis, while the small
    open circles below $0.6\,\hmpc$ are left unused by our model fit due to the
    potential impact from fibre collision. The vertical shaded region represents
    the distance scales that are below the maximum fibre-collided distance scale
    at the maximum redshift of the sample~($z{=}0.3$). Thick solid curve is the
    posterior mean prediction from our MCMC analysis, with the thin bundle of
    curves of the same colour showing the predictions from 100 random steps
    along the MCMC chain. The two other solid curves with different colours~(but
    consistent across the three panels) indicate the posterior mean predictions
    for the other two stellar mass subsamples. Dotted and dashed lines indicate
    the contributions from the central-satellite pairs and satellite-satellite
    pairs within the same halo, respectively. Dot-dashed line indicates the
    contribution from the galaxy pairs between different haloes.}
\label{fig:lowzwps}
\end{center}
\end{figure*}
\begin{figure*}
\begin{center}
    \includegraphics[width=0.96\textwidth]{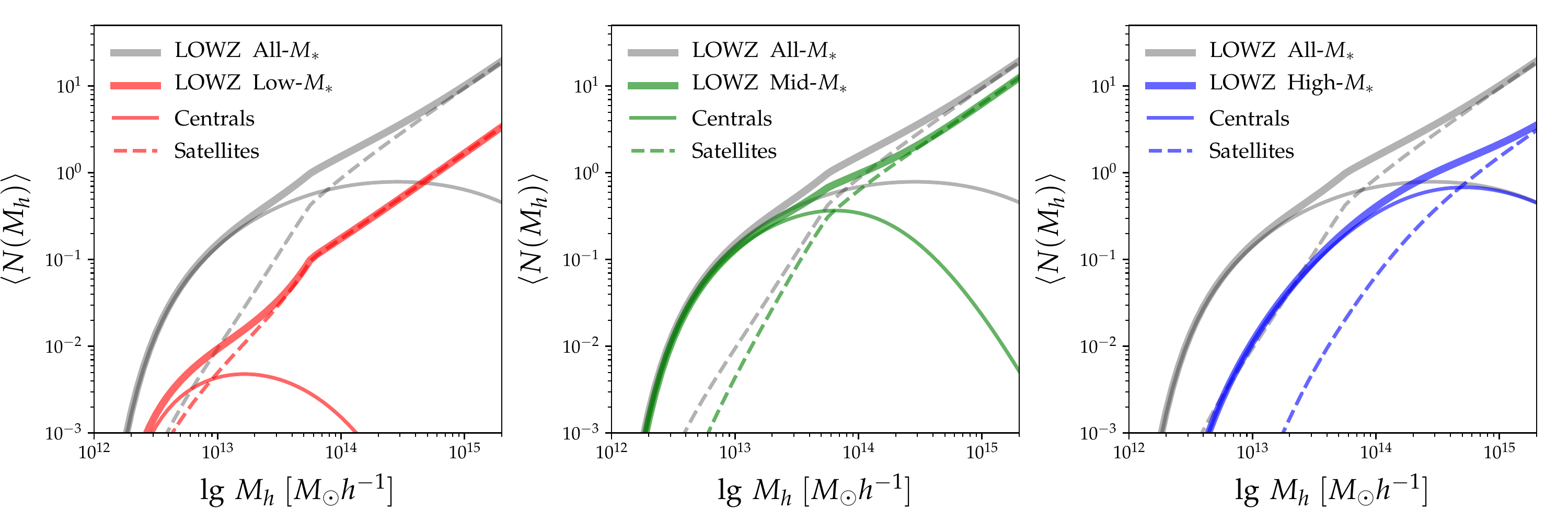} \caption{HODs
    predicted by the best-fitting model for the Low-$M_*$~(left),
    Mid-$M_*$~(middle), High-$M_*$~(right) subsamples in the \lowz{} sample. In
    each panel, the red, green, or blue curves describe the set of HODs for the stellar
    mass subsample, while the gray curves are for the entire \lowz{}
    sample~(same across all three panels). For each set of HODs, thin solid and
    thin dashed curves indicate the HODs of the central and satellite galaxies,
    respectively, and the sum of the two are shown as the thick solid curve.}
\label{fig:lowzhods}
\end{center}
\end{figure*}

At fixed stellar mass, the central galaxies reside in haloes within a narrow
range of halo mass because of the small scatter in the SHMR~(${\sim}0.2$ dex).
Therefore, it is unsurprising that the central selection function is insensitive
to halo mass. However, the host haloes of the satellites usually have a much
larger spread in halo mass, and the quenched fraction of satellites increases
significantly with halo mass. Thus, we anticipate an extra halo mass dependence
in the satellite selection function of the BOSS samples,
\begin{equation}
\fdet^{\mathrm{sat}}(\ms) = \frac{f_{\mathrm{max}}}{2} \left( 1 + \erf \left[ \left(\lg M_* -
    \lg M_{*}^{s}(\mh)\right) / \sigma^{s}(\mh)\right] \right),
    \label{eqn:fdetsat}
\end{equation}
where
\begin{equation}
    \lg M_{*}^{s}(\mh) = \left\{
    \begin{array}{ll}
        \lg M_{*}^{s} + \mu (\lg\mh - \lg M_{h}^s) ,&\mbox{ $\mh<M_{h}^s$} \\
        \lg M_{*}^{s} ,&\mbox{ $\mh\geq M_{h}^s$,}
    \end{array} \right.
    \label{eqn:mu}
\end{equation}
and
\begin{equation}
    \sigma^{s}(\mh) = \left\{
    \begin{array}{ll}
        \sigma^{s} + \nu (\lg\mh - \lg M_{h}^s) ,&\mbox{ $\mh<M_{h}^s$} \\
        \sigma^{s} ,&\mbox{ $\mh\geq M_{h}^s$,}
    \end{array} \right.
    \label{eqn:nu}
\end{equation}
respectively. In Equations~\ref{eqn:mu} and~\ref{eqn:nu}, $M_{h}^s$ is a
characteristic halo mass beyond which the satellite selection function depends
only on stellar mass, and $\mu$ and $\nu$ dictate the variation of the
characteristic stellar mass and transition width with halo mass below $M_{h}^s$,
respectively.  We assume the same $f_{\mathrm{max}}$ for centrals and satellites
during the fit as it is likely independent of galaxy properties.  Therefore, the
satellite detection fraction at fixed halo mass has a similar form compared to
the central detection fraction~(without the high mass cut-off), but both the
characteristic stellar mass and the transition width vary with halo mass below
$M_{h}^s$. The halo-mass independence above $M_{h}^s$ is intended to mimic the
fact that the satellite colours have converged to the red-sequence in the most
massive systems. Below $M_{h}^s$, non-zero values of $\mu$ and $\nu$ would
select galaxies with different stellar mass distributions into the BOSS sample.

\begin{figure*}
\begin{center}
    \includegraphics[width=0.96\textwidth]{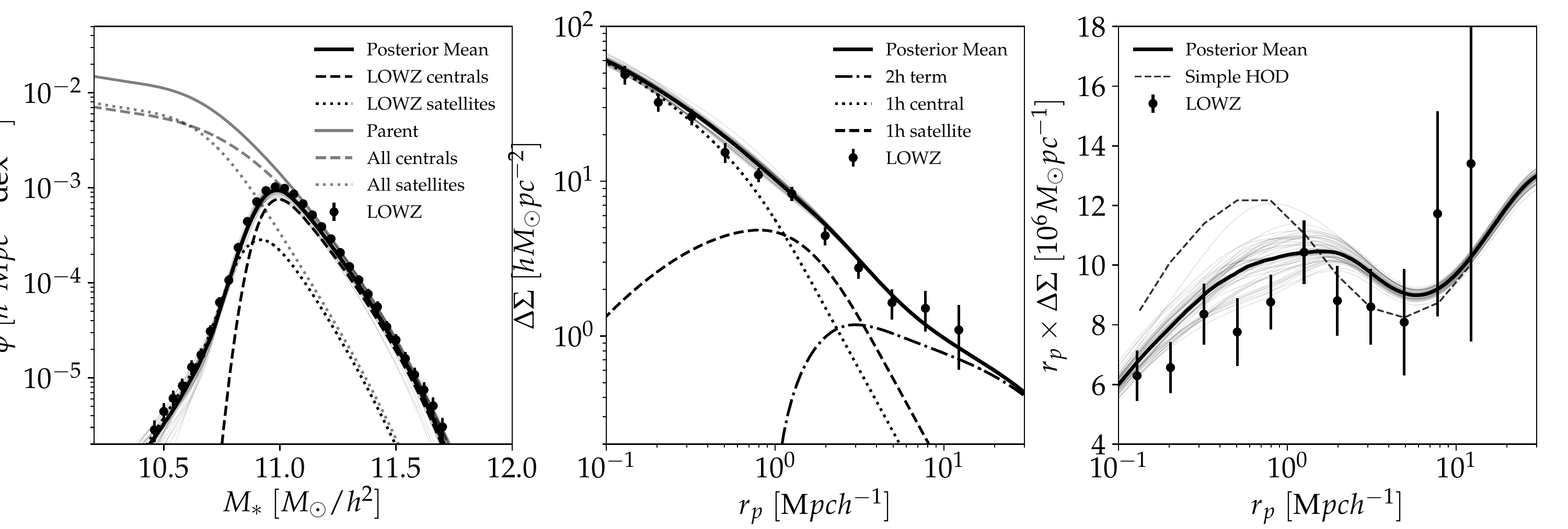} \caption{{\it
    Left}: Comparison between the observed~(data points with errorbars) and
    posterior mean predicted~(thick solid curve) SMFs for the \lowz{} sample.
    The bundle of thin solid curves surrounding the thick solid curve indicates
    the predictions from 100 random steps along the MCMC chain, while the dashed
    and dotted black curves represent the contributions from the central and
    satellite galaxies, respectively. Gray solid, dashed, and dotted curves are
    the SMFs of the stellar mass-complete sample predicted by the posterior mean
    model. {\it Middle}: Comparison between the observed~(data points with
    errorbars) and posterior mean predicted~(thick solid curve) g-g lensing
    signals. The thin bundle of curves are the predictions from 100 random MCMC
    steps.  Dotted, dashed, and dot-dashed curves are the contributions from the
    one-halo central, one-halo satellite, and two-halo lensing terms,
    respectively. {\it Right}: Similar to the middle panel, but with the y-axis
    replaced by $r_p{\times}\ds$. Dashed curve shows the prediction from the
    model derived from clustering in \citet{Lange2019}. The posterior mean
    prediction from our model successfully reproduces the observed SMF and g-g
    lensing signal on all scales.}
\label{fig:lowzsmfds}
\end{center}
\end{figure*}

Figure~\ref{fig:fdet} illustrates the 2D satellite selection function~(top
panel) and the 1D satellite~(coloured curves in the bottom panel) vs. central
selection functions~(gray dashed curve in the bottom panel), respectively,
predicted by the best-fitting model for the \lowz{} sample~(as will be described
later in \S~\ref{sec:lowz}). The 2D satellite selection map indicates that the
selection function depends only on stellar mass at $\mh{>}13.76$, but diverges
into a wider transition and higher stellar mass threshold with decreasing halo
mass. Compared to the central selection function~(gray dashed), the 1D satellite
selection functions have similarly sharp transitions and characteristic stellar
mass at high $\mh$, but selects significant lower fraction of galaxies at high
stellar mass from low mass haloes.

To summarise the selection function prescription, we have eight model
parameters. Among them $\{\lg M_{*}^c, \sigma^c\}$ describe the selection
function of centrals galaxies, $\{\lg M_{h}^s, \lg M_{*}^s, \sigma^s, \mu, \nu
\}$ describe the 2D selection function of satellite galaxies, and
$f_{\mathrm{max}}$ is the maximum detection fraction of the sample.

Finally, combining the HOD~(\S~\ref{subsec:hod}) and selection
function~(\S~\ref{subsec:selection}) prescriptions, our model has 19 parameters.
Note that we have employed the full HOD prescription from \citet{Zu2015}, which
was designed to fit all the galaxies above $\lg M_*{=}8.5$, while our minimum
stellar mass is approximately two orders of magnitudes higher at
$\lg M_*{=}10.5$. As a result, most of the parameters that describe the
low-mass end behaviours will be prior-dominated. Furthermore, the investigation
of redMaPPer BCGs in ~\S~\ref{subsec:bcg} has greatly reduced the number of
parameters for the central galaxy selection modelling compared to the
satellites.

\subsection{Gaussian Likelihood Model}
\label{subsec:like}

Equipped with the capability of predicting the SMF $\phi(\ms)$, projected
auto-correlation function $w_p(r_p)$, and g-g lensing signals $\ds(r_p)$ for
each of the four BOSS samples, we can infer the posterior probability
distribution of the model parameters from fitting the three observables within a
Bayesian framework assuming a Gaussian likelihood model.

We include two components in the error matrices of the observed SMFs, one is the
Poisson errors from number counting, the other is an extra term representing the
sample variance and uncertainties in the comoving volume calculation due to the
uncertainties in cosmology and effective area. For the second term, we add an
additional $10$ per cent error in the diagonal term with a $50$ per cent
covariance in the off-diagonal terms. The error matrices of $w_p$ is directly
estimated from the data using the jackknife re-sampling technique, by dividing
the LOWZ/CMASS northern sky coverage into $200$ patches of the similar area. We
adopt the same error matrices of the g-g lensing signals from ~\citet{Lange2019}
and ~\citet{Leauthaud2017} for LOWZ and CMASS galaxies, respectively. When
combining the three observables into one data vector, we ignore the weak
covariance among the three types of observables, and among the different stellar
mass subsamples of the same redshift bin.

We model the combinatorial vector $\mathbf{x}$ of $\phi$, $w_p$, and $\ds$ as a
multivariate Gaussian, which is fully specified by its mean~($\bar{\mathbf{x}}$)
and covariance matrix~($\mathbf{C}$). The Gaussian likelihood is thus
\begin{equation}
    \mathcal{L}(\mathbf{x} | \boldsymbol{\theta}) =
    |\mathbf{C}|^{-1/2}\exp\left(-\frac{(\mathbf{x}-\bar{\mathbf{x}})^T\mathbf{C}^{-1}(\mathbf{x}-\bar{\mathbf{x}})}{2}\right),
\label{eqn:gauloglike}
\end{equation}
where
\begin{multline}
    \boldsymbol{\theta} \equiv
    \{\lg\mh^1,\lg\ms^0,\beta,\delta,\gamma,B_{\mathrm{sat}},\beta_{\mathrm{sat}},
    \alpha_{\mathrm{sat}}, \sig, \eta, f_c,\\ \lg M_{*}^c, \sigma^c, \lg M_{h}^s, \lg M_{*}^s, \sigma^s, \mu, \nu, f_{\mathrm{max}} \}.
\end{multline}

To focus our constraint on the stellar mass range above $\lg\ms{>}10.5$, we
place priors on both the slope and scatter of the SHMR at the low-mass end.  In
particular, we limit the predicted mean stellar mass at halo mass
$10^{11}\,\hmsol$ to be between $10^8\hhmsol$ and $10^9\hhmsol$, informed by the
study of the SHMR evolution with redshift in~\citet{Yang2012}. We also place a
Gaussian prior on $\sig{\sim}\,\mathcal{N}(0.50,\,0.04^{2})$, informed by the
constraints from \citet{Zu2015}. The slope prior is not directly applied on any
of the parameters because our SHMR parameterization does not allow a clean
separation of the low and high-mass end slopes, which is however possible in
other parameterisations\citep[e.g.,][]{Yang2012}. Finally we place a Gaussian
prior on $f_{\mathrm{max}}{\sim}\,\mathcal{N}(0.86,\,0.05^{2})$, informed by our
direct measurement from \S~\ref{subsec:bcg}.

For each redshift bin, the joint posterior distribution of the parameters is
derived using the Markov Chain Monte Carlo~(MCMC)
algorithm~\texttt{emcee}~\citep{Foreman-Mackey2013}, where an affine-invariant
ensemble sampler is utilised to fully explore the parameter space. For each MCMC
chain, we perform $40,0000$ iterations, $100,000$ of which belong to the burn-in
period for adaptively tuning the steps. To eliminate the residual correlation
between adjacent iterations, we further thin the chain by a factor of $10$ to
obtain our final results. The 68 per cent confidence regions of the 1D posterior
constraints are listed in Table~\ref{tab:constraints}, with one column for each
of four redshift bins.

\section{Results from LOWZ}
\label{sec:lowz}

We highlight the results from our joint analysis of the $\Phi$, $w_p$, and $\ds$
for the \lowz{} sample in this Section, starting by examining the model fits to
the $w_p$ measurements of the three stellar mass bins below. Compared to
\citet{Lange2019} analysis, we employ the same g-g lensing signals measured by
their work, but divide the LOWZ galaxies into three stellar mass subsamples,
each of which covers a distinctive portion of the observed SMF, rather than two
(arbitrary) subsamples of equal logarithmic bin width.  The deliberate stellar
mass binning helps elicit the important information on the dependence of
large-scale bias on stellar mass, and more important, satellite fraction.

Figure~\ref{fig:lowzwps} compares the measured projected auto-correlation
functions to predictions from our posterior mean model for the three
stellar-mass bins in the \lowz{} sample~(stellar mass increases from left to
right).  In each panel, circles with errorbars show the $w_p$ measurements for
the particular stellar mass bin, with large solid circles indicating the data
points used for the model fit and small open ones the data points unused due to
fibre collision. The vertical shaded region on the left of each panel indicates
the scales below the maximum comoving distance that the fibre radius corresponds
to at $z{=}0.3$, which is well below the $0.6\,\hmpc$ minimum fitting scale that
separates the large solid and small open circles. The posterior mean prediction
from our model is indicated by the thick solid curve, with the bundle of thin
curves of the same colour indicating the predictions from 100 random steps along
the MCMC chain. For the sake of straight comparison of the three stellar mass
bins, we also show the respective posterior mean predictions for the other two
stellar mass bins as the two solid lines of different colours~(but consistent
with the colours in their respective panels). Under the solid curves, the dotted
and dashed curves indicate the contributions from the one-halo central-satellite
and the one-halo satellite-satellite terms to the overall predicted signal,
respectively, while the dot-dashed curve indicates the two-halo term.

\begin{figure*}
\begin{center}
    \includegraphics[width=0.96\textwidth]{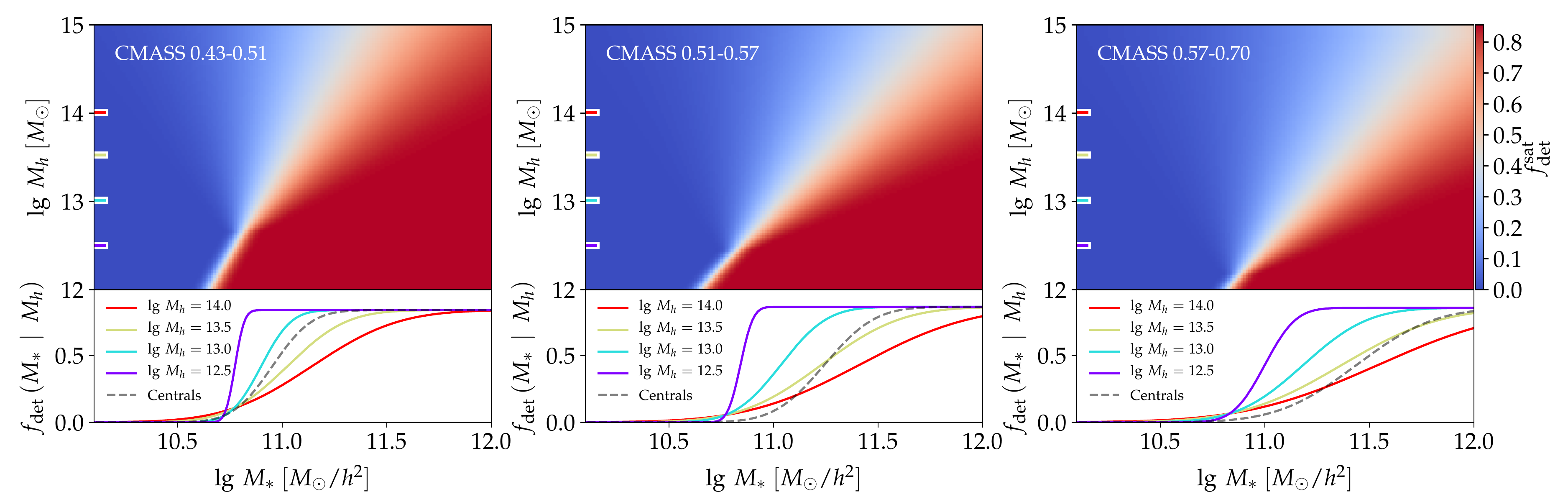}
    \caption{Similar to Figure~\ref{fig:fdet}, but for \cmassa~(left),
    \cmassb~(middle), and \cmassc~(right), respectively. The selection function in
    CMASS preferentially observes satellite galaxies from low-mass haloes,
    compared to LOWZ which selects more satellites from high-mass haloes.}
\label{fig:fdetcmass}
\end{center}
\end{figure*}

On large scales, the $w_p$ signal has a non-monotonic dependence on stellar
mass, with the Mid-$M_*$ subsample~(green) showing the lowest galaxy bias than
the Low-$M_*$ and High-$M_*$ subsamples which show comparable clustering
amplitudes to each other. Such non-monotonic behavior of galaxy bias is well
described by the model predictions. The shape of the measured $w_p$ of the
Low-$M_*$ bin shows a deviation from that of a biased matter clustering on
scales larger than $20\hmpc$, but is still consistent with model prediction
within the large errorbars. Meanwhile, the measured $w_p$ of the High-$M_*$ bin
exhibits a flattening feature at $r_p{\simeq}2\hmpc$, followed an enhancement
compared to the prediction on scales between $3{-}5\hmpc$. Since the galaxies in
the high mass bin are dominated by the central galaxies of massive haloes, this
feature is likely associated with the galaxy kinematics and distribution in the
infall region, which may require specially-tailored models like those
in~\citet{Zu2013}. On small scales, the clustering is subject to the fibre
collision effect, but the model predictions agree reasonably well with the small
open circles that are above the maximum fibre radius~(i.e., outside of the gray
shaded region). This agreement, combined the fact that we do not see a sudden
decrease in the clustering amplitude, suggests that the impact from fibre
collision is weak in the \lowz{} sample.

To understand the behaviors of the measured and predicted $w_p$ signals, we
examine the HOD of each of three subsamples in
Figure~\ref{fig:lowzhods}~(stellar mass increases from left to right).  In each
panel, we show the HODs of the entire \lowz{} sample in gray as the reference
HOD, and the results of the stellar mass subsample in one of red~(Low-$M_*$),
green~(Mid-$M_*$), or blue~(High-$M_*$) colours. For each sample, we show the
HODs of the central and satellite galaxies as thin solid and thin dashed curves,
respectively, and the sum of the two in thick solid curve. As expected, the
Low-$M_*$ subsample is dominated by satellite galaxies in massive haloes, while
the High-$M_*$ subsample consists mainly of central galaxies in haloes of
similar mass. Therefore, although the average stellar mass of those two
subsamples are different by ${\sim}0.5$ dex, they exhibit almost the same
large-scale galaxy bias. The satellite HOD of the Low-$M_*$ subsample~(red
dashed curve in left panel) shows a kink at $\lg\mh{=}13.76$, where the
satellite selection function starts to depend on halo mass, thereby showing a
steeper slope than at the high-$\mh$ end~(${\sim}1$). Clearly, this kink is
directly related to the deflection of the half-$f_{\mathrm{max}}$ track~(delineated
by the white colour) on the satellite detection map in Figure~\ref{fig:fdet}.

In the middle panel of Figure~\ref{fig:lowzhods}, the HOD of the Mid-$M_*$
subsample is dominated by the central galaxies in group-size haloes, hence the
lower large-scale bias than the other two, which reflect the bias of
cluster-size haloes. We emphasize again that, since the HOD of the total \lowz{}
sample is dominated by the Mid-$M_*$ subsample at the peak of the observed SMF,
fitting to the $w_p$ measurements of the total sample alone would not provide
the key information on the halo mass dependence of the satellite selection
function revealed by the Low-$M_*$ and High-$M_*$ subsamples.

Given that the HODs displayed in Figure~\ref{fig:lowzhods} provide excellent
description to the nontrivial stellar mass dependence of galaxy clustering on
both small and large scales, it is intriguing whether it could also reproduce
the observed SMF, and most importantly, the g-g lensing signals. We first
compare the observed and posterior mean SMFs in the left panel of
Figure~\ref{fig:lowzsmfds}, where data points with errorbars are the measured
SMF and thick solid curve is our posterior mean prediction, with the thin bundle
of curves showing predictions from 100 random steps along the MCMC chain.
Underneath the thick solid curve we decompose the predicted SMF into
contributions from the central~(black dashed) and satellite~(black dotted)
galaxies. The gray solid, dashed, and dotted curves are the predicted total,
central, and satellite SMFs of the parent \lowz{} sample~(i.e., 100 per cent
completeness). The posterior mean SMF provides an excellent description of the
observed stellar mass distribution of \lowz{} galaxies, including the slope
change at the low mass end. As expected from our intuition in
\S~\ref{subsec:boss}, the central SMF has a sharp cutoff at $\lg\ms{=}10.94$,
vacating the low stellar mass portion to satellite galaxies.  The slope change
at $\lg\ms{=}10.7$ is the direct consequence of preferentially selecting
galaxies at the low stellar mass tail from massive halos. One useful test of our
model is to compare the predicted parent SMF~(thin solid curve) to those
inferred from multi-band deep imaging data by, e.g, \citet{Leauthaud2016};
Unfortunately the large discrepancy in the stellar mass estimates from
spectroscopy vs. imaging makes such direct comparison difficult~(see figure 15
in their paper).

\begin{figure*}
\begin{center}
    \includegraphics[width=0.96\textwidth]{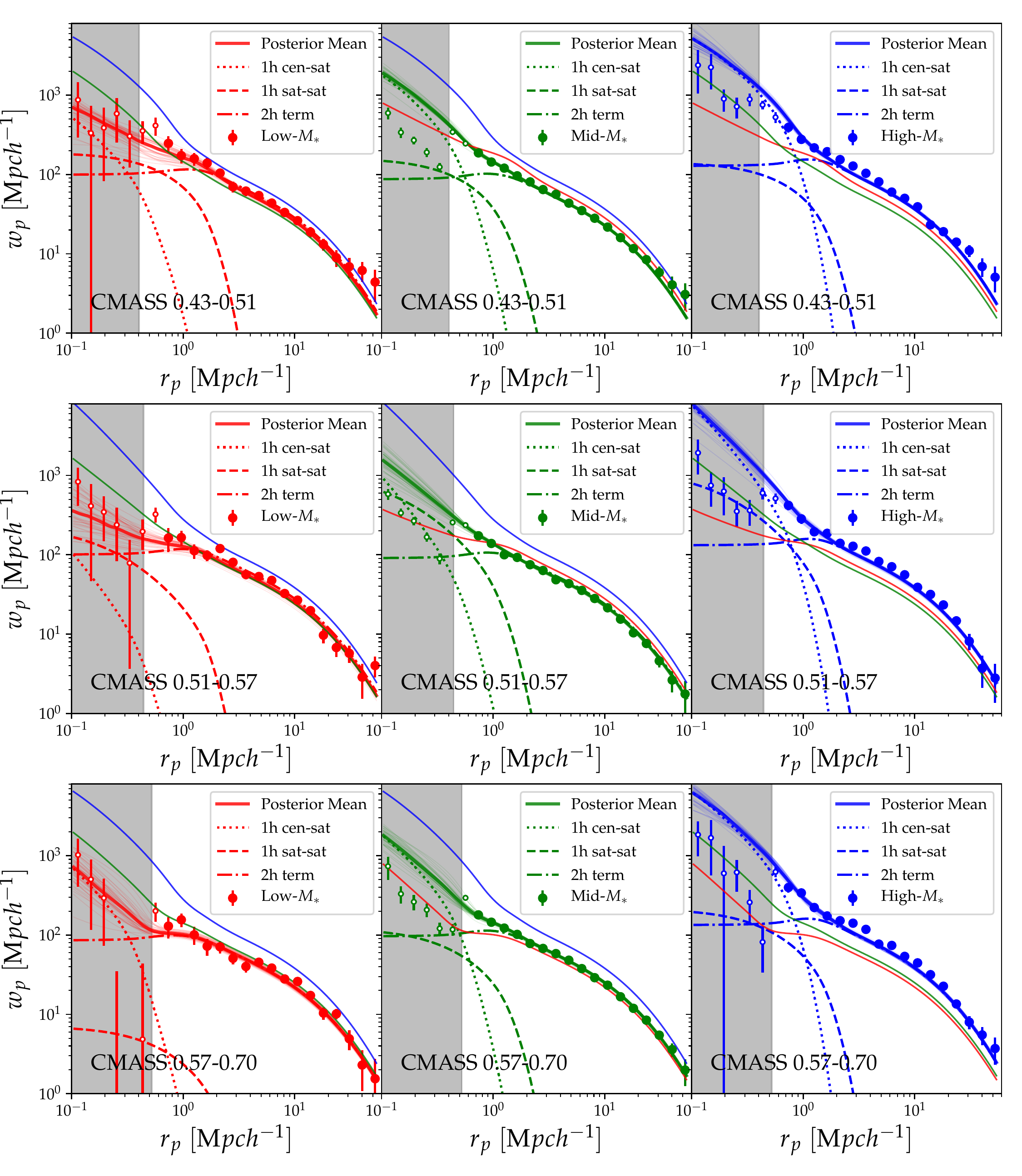} \caption{
        Similar to Figure~\ref{fig:lowzwps}, but for the \cmassa{}~(top row),
        \cmassb{}~(center row), and
        \cmassc{}~(bottom row) subsamples, respectively.  }
\label{fig:cmasswps}
\end{center}
\end{figure*}

Finally, we examine the g-g lensing signals in the middle and right panels of
Figure~\ref{fig:lowzhods}. In both panels, data points with errorbars are the
measurements for the entire \lowz{} sample, and thick solid curves are the
posterior mean prediction from our model, with the thin bundle of curves
indicating the individual predictions by 100 random steps along the MCMC chain.
We multiply $\ds$ by $r_p$ in the right panel to facilitate signal comparison
across small and large scales. In the middle panel of Figure~\ref{fig:lowzhods},
we also show the decomposition into host halo lensing signal around
centrals~(dotted), host halo lensing signal around satellites~(dashed), and the
large-scale lensing signal from other neighbouring haloes~(dot-dashed). The
posterior model prediction provides an excellent fit to the g-g lensing
measurements on all scales, without any strong lensing discrepancy on small
scales. In the right panel of Figure~\ref{fig:lowzhods}, we additionally show
the prediction from the best-fitting conditional stellar mass function model of
\citet{Lange2019}~(thin dashed curve), which does a slightly better job than our
model on scales larger than $2\hmpc$. However, the dashed curve significantly
over-predicts the lensing signal on scales below $1\hmpc$ compared to the data
points and to our posterior mean prediction.

To summarise, our best-fitting model successfully reproduces the observed SMF,
galaxy clustering, and g-g lensing signals simultaneously at the Planck
cosmology. The carefully binning in the stellar mass allows us a better
constraint of satellite fraction $f_{\mathrm{sat}}$, which is signifiantly
non-zero at the low stellar mass end. Meanwhile, the model allows the selection
function of satellite galaxies to have an extra halo mass dependence, which
breaks the lockstep between the small-scale lensing and large-scale galaxy bias
that are only valid in the limit of $f_{\mathrm{sat}}{\rightarrow}0$.

\section{Results from CMASS}
\label{sec:cmass}

In the previous Section we demonstrate that our model is capable of fully
resolving the discrepancy between the large-scale clustering and small-scale
lensing of the \lowz{} sample at Planck cosmology. We now move on to the three
CMASS samples in which the original ``lensing is low'' discrepancy was
discovered by \citet{Leauthaud2017}.  We apply our HOD analysis to the
\cmassa{}, \cmassb{}, and \cmassc{} samples separately with the same set-up as
in \lowz{}. Since the constraints from the three samples are qualitatively
similar, we will describe their results simultaneously below.

Figure~\ref{fig:fdetcmass} shows the galaxy selection functions predicted by the
best-fitting models for \cmassa~(left), \cmassb~(middle), \cmassc~(right),
respectively. The format of each panel is the same as Figure~\ref{fig:fdet}.
Compared to the results from \lowz{}, the CMASS satellite selection functions
inferred by the model exhibit a preference of selecting satellite galaxies from
the {\it low-mass} haloes, rather than {\it high-mass} haloes as shown in
Figure~\ref{fig:fdet}. This change of preference is likely caused by the
combination of two factors, one observational and the other physical. Firstly,
the CMASS colour selection includes more galaxies that are in the blue cloud,
while the LOWZ selection follows more faithfully the traditional LRG colour
cuts~\citep{Eisenstein2001}. Secondly, galaxies in massive haloes are in general
bluer at higher redshifts~\citep{Cooper2007, Hansen2009, Nishizawa2018}.  In
addition, Figure~\ref{fig:fdetcmass} shows that this preference becomes even
stronger for CMASS galaxies at higher redshifts, bringing relatively more
galaxies from the low-mass haloes into the sample. This redshift dependence is
consistent with the findings in \citet{Montero-Dorta2016}, who estimated that
the fraction of intrinsically blue galaxies in CMASS increases from ${\sim}36$
per cent at $z{=}0.5$ to ${\sim}46$ per cent at $z{=}0.7$ --- satellite galaxies
from lower-mass haloes are generally bluer.

\begin{figure*}
\begin{center}
    \includegraphics[width=0.96\textwidth]{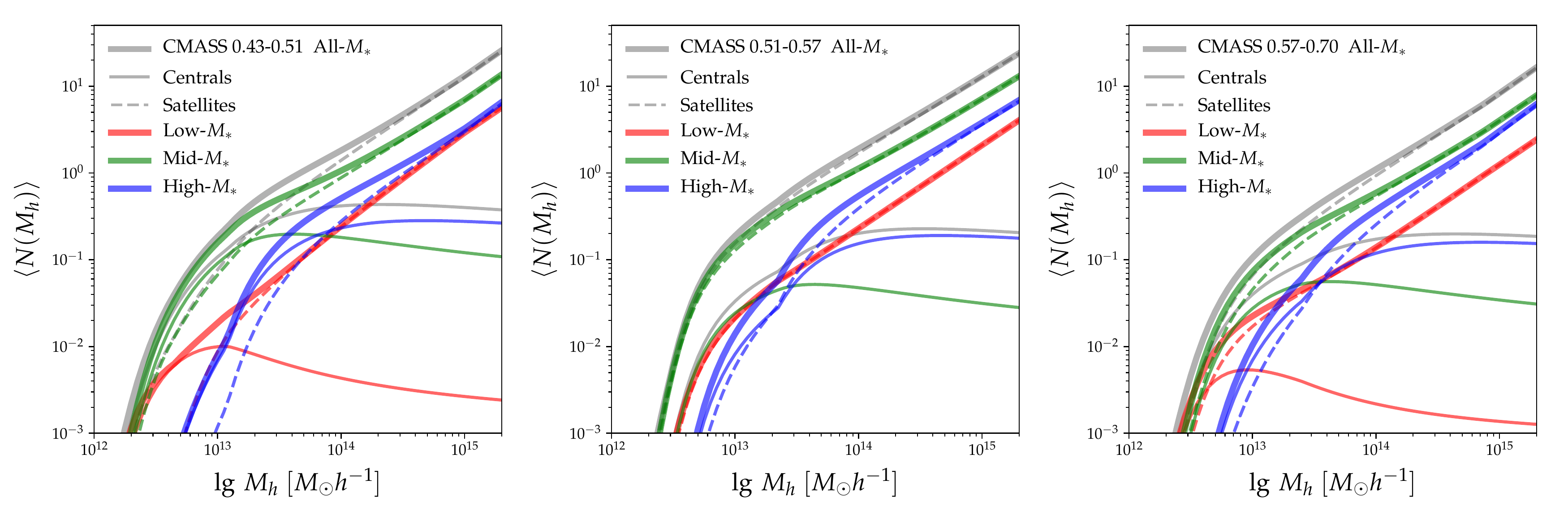} \caption{Similar
    to Figure~\ref{fig:lowzhods}, but for the \cmassa{}~(left),
	\cmassb{}~(middle), and \cmassc{}~(right) subsamples, respectively.
	Instead of showing the three stellar mass subsamples in separate panels
	as in
    Figure~\ref{fig:lowzhods}, we display them in the same panel for each
    redshift bin, with Low-$M_*$, Mid-$M_*$, and High-$M_*$ HODs shown in red, green
    and blue, respectively. Gray curves are the HODs of the overall
    sample. }
\label{fig:cmasshods}
\end{center}
\end{figure*}

Figure~\ref{fig:cmasswps} compares the projected auto-correlation functions
between the measurements and posterior mean predictions, for the \cmassa{}~(top
row),  \cmassb{}~(center row),  and \cmassc{}~(bottom row), respectively. The format
of each row is the same as Figure~\ref{fig:lowzwps}, with the Low-$M_*$,
Mid-$M_*$, and High-$M_*$ results shown by the left, middle, and right panels,
respectively. The most important difference among the three stellar mass
subsamples is the stellar mass trend of the $w_p$ signal on large scales. While
the Mid-$M_*$ and High-$M_*$ signals show consistent trends across different
redshifts, the relative bias of Low-$M_*$ galaxies decreases as a function of
redshift. However, even in the \cmassa{} subsample where the bias of the
Low-$M_*$ galaxies is the highest, it is still substantially lower than that of
the High-$M_*$ ones, unlike the case in \lowz{} where they are comparable.  On
small scales, the impact of fibre collision is clearly seen among the small open
circles, which always experience a sudden drop entering the gray shaded region,
i.e., the fiber-collided scales. Beyond the shaded region, However, the small
open circles are largely consistent with our predictions, despite that those
data points were never used for the model fit. Interestingly, for the Low-$M_*$
subsamples, the model predictions agree well with the measured $w_p$ even inside
the fibre-collided regime. One possibility is that the Low-$M_*$ subsamples
consist mostly satellite galaxies without a dominant central in the nearby sky,
hence a lower fibre collision probability.

\begin{figure*}
\begin{center}
    \includegraphics[width=0.96\textwidth]{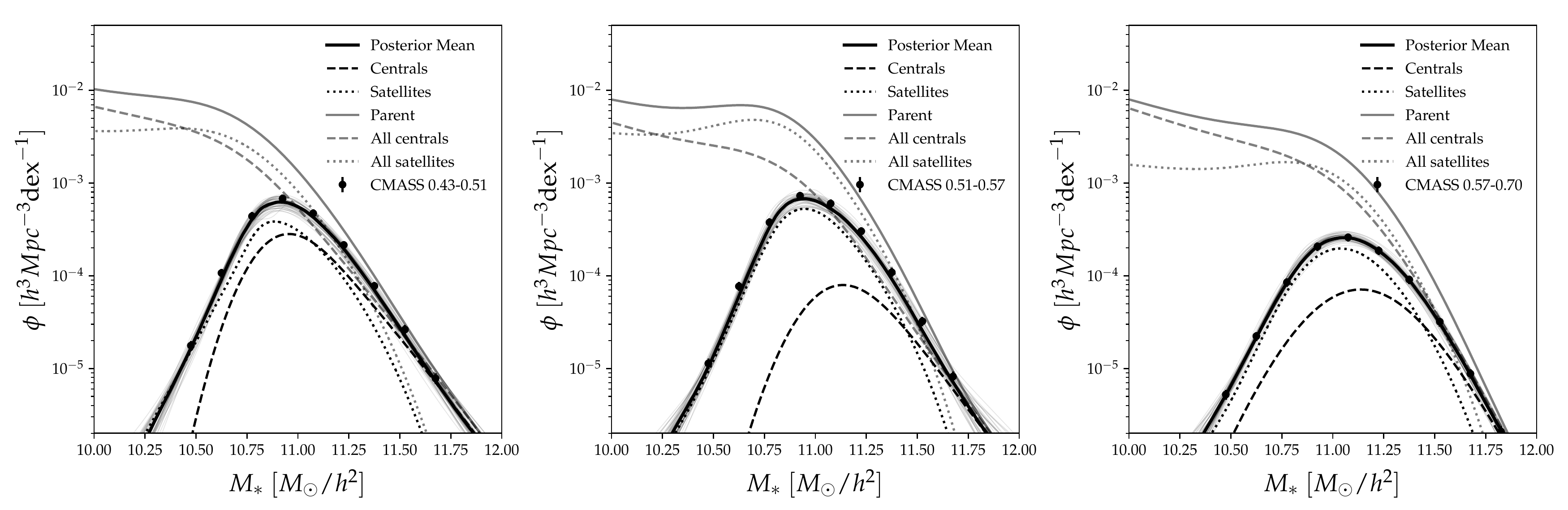} \caption{Similar
    to the left panel of Figure~\ref{fig:lowzsmfds}, but for the
    \cmassa{}~(left),
	\cmassb{}~(middle), and \cmassc{}~(right) subsamples, respectively.
    }
\label{fig:cmasssmfs}
\end{center}
\end{figure*}

\begin{figure*}
\begin{center}
    \includegraphics[width=0.96\textwidth]{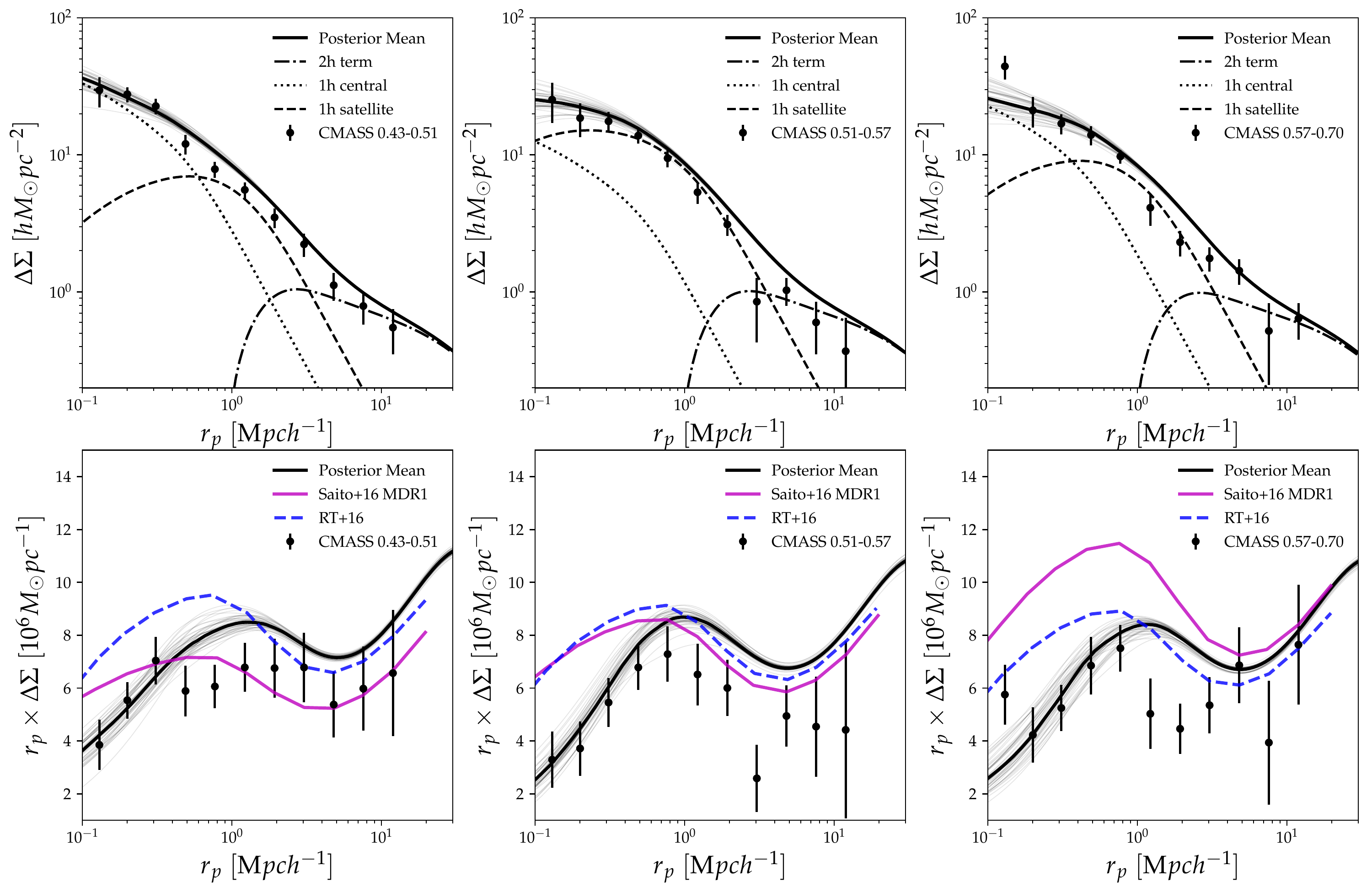} \caption{{\it
    Top}: Comparison between the measured g-g lensing signals~(data points with
    errorbars) and the posterior mean predictions~(thick solid curves surrounded
    by thin bundle of gray curves) from our model, for the \cmassa{}~(left),
    \cmassb{}~(middle), and \cmassc{}~(right), respectively. Each thick solid
    curve is decomposed into contributions from the one-halo central~(dotted),
    one-halo satellite~(dashed), and two-halo terms, respectively. {\it Bottom}:
    Similar to the top panels, but with the y-axis changed to $r_p\times\ds$. In
    each panel, magenta solid and blue dashed curves are predictions that
    exhibit the original ``lensing is low'' discrepancy, from \citet{Saito2016}
    and \citet{Rodriguez-Torres2016}, respectively. Our posterior model
    predictions provide an excellent description to the g-g lensing measurements
    on small scales below~$1\,\hmpc$, and a comparable fit to the large scales
    compared to the other two models.}
\label{fig:cmassdss}
\end{center}
\end{figure*}

The behaviours of the measured and predicted $w_p$ signals can be understood by
examining their HODs in Figure~\ref{fig:cmasshods}. The format of the Figure is
similar to that of Figure~\ref{fig:lowzhods}, but we display the results for the
three stellar mass subsamples in one panel instead of showing them separately.
Overall, the stellar mass dependence of the HODs at each redshift is similar to
one another. In general, the satellite HODs of the Low-$M_*$~(red dashed curves)
subsamples exhibit an enhancement at the low halo mass due to the preferential
selection seen in Figure~\ref{fig:fdetcmass}, which also helps bring down the
relative galaxy bias of the CMASS Low-$M_*$ subsamples compared to that in LOWZ.
Intriguingly, the Low-$M_*$ subsample of the \cmassb{} bin barely has any
central galaxies. This is likely driven by the relatively low lensing amplitude
on small scales~(middle panels of Figure~\ref{fig:cmassdss}), as its impact on
$w_p$ is negligible.

We compare the observed SMFs to those predicted by our inferred HODs in
Figure~\ref{fig:cmasssmfs}. The format of each panel is the same as in the left
panel of Figure~\ref{fig:lowzsmfds}. Unlike the \lowz{} sample, there does not
appear any feature on the observed CMASS SMFs~(data points with errorbars), so
the SMFs are less constraining compared to that in Figure~\ref{fig:lowzsmfds}.
Nevertheless, we obtain excellent fits to the data with the same HODs that
reproduce the clustering measurements. Therefore, the best-fitting HODs provide
an excellent description of the joint redshift and stellar mass dependences of
the clustering and abundance for CMASS galaxies.

We now examine the g-g lensing signals predicted by the best-fitting HODs, and
compare if they remain systematically higher than observations on small scales.
Figure~\ref{fig:cmassdss} shows a comprehensive comparison between the observed
g-g lensing signals~(data points withe errorbars) and our posterior mean
predictions~(thick solid curves surrounded by a bundle of thin solid curves),
for the \cmassa{}~(left panels), \cmassb{}~(middle panels), and \cmassc{}~(right
panels), respectively. In the top panels, we decompose the predictions into
one-halo central~(dotted), one-halo~satellite~(dashed) and two-halo~(dot-dashed)
terms. Meanwhile in the bottom panels, we highlight the comparison by showing
$r_p\times\ds$ as the y-axis, and add two extra curves predicted by the
\citet{Saito2016}~(magenta solid) and \citet{Rodriguez-Torres2016}~(blue dashed)
analyses, both of which exhibit very strong lensing discrepancy on small scales.
The \citet{Saito2016} predictions show similar large-scale lensing amplitudes to
our curves, but on small scales are higher than
the measurements by ${\sim}40\%$ in all three redshift bins. The
\citet{Rodriguez-Torres2016} results show better agreement with the data for the
\cmassa{} sample on scales above $0.3\hmpc$ compared to our curve, but
significantly over-predict the signal on all scales for the other two samples,
reaching a discrepancy as large as $50\%$ for the \cmassc{} sample on small scales.

On scales above $1\hmpc$, our predictions are in general higher than the
observations by ${\sim}30$ per cent, except for the High-$M_*$ subsample.  This
is slightly concerning --- the large-scale lensing signal is fully determined by
the galaxy bias at fixed cosmology, which is then tightly constrained by the
projected clustering shown in Figure~\ref{fig:cmasswps}.  Therefore, within our
model framework at best-fitting Planck cosmology, it is very difficult to lower
the large-scale lensing amplitude to be better matched to the measurements. In
terms of uncertainties in the lensing signals, \citet{Leauthaud2017} did a
comprehensive analysis of the possible sources of systematic errors, and
concluded that the total fractional systematic error on $\ds$ is $5{-}10$ per
cent, much smaller than the large-scale discrepancy we see in
Figure~\ref{fig:cmassdss}.

On scales below $1\hmpc$, however, our predicted signals are in excellent
agreement with the measurements, without showing any symptoms of a low lensing
amplitude. This is very encouraging, indicating that the CMASS lensing
discrepancy on small scales is also resolved, similar to that in our \lowz{}
analysis.  As pointed out by \citet{Leauthaud2017} and emphasized in
\S~\ref{sec:intro}, the essence of the ``lensing is low'' discrepancy lies in
the one-halo regime, where the observed small-scale lensing signal is lower than
that inferred from the large-scale galaxy bias. Therefore, given that our model
has provided excellent fits to the comprehensive galaxy clustering in
Figure~\ref{fig:cmasswps} and the small-scale g-g lensing in
Figure~\ref{fig:cmasswps}, the original ``lensing is low'' discrepancy revealed
by the CMASS galaxies is also reasonably resolved by our HOD model via the
incorporation of a halo mass dependence in satellite selection.

\section{Conclusion}
\label{sec:conc}

In this paper, we have applied an analytic model based on the \ihod{} framework
of \citet{Zu2015, Zu2016}, to the observed stellar mass functions $\phi$,
projected auto-correlation functions $w_p$, and g-g lensing signals $\ds$ of
galaxies in the BOSS LOWZ and CMASS samples. Our main conclusion is that, it is
viable for the combination of our standard HOD model and the best-fitting Planck
$\lcdm$ cosmology to resolve the discrepancy between the large-scale galaxy bias
and the small-scale g-g lensing signals, originally discovered by
\citet{Leauthaud2017} for CMASS and subsequently for LOWZ~\citep{Lange2019}.

We directly measure the LOWZ selection function of the central galaxies of the
\redmapper{} clusters, finding that the central selection function is largely
independent of halo mass, at least in the cluster mass range. For the satellite
galaxies, we developed a 2D galaxy selection function that allows the satellite
detection fraction to decline from high completeness at high $M_*$ to zero at
low $M_*$, but with the characteristic stellar mass and width of the decline
both dependent on the mass of the host haloes. Additionally, we carefully divide
the galaxies of each redshift bin into three different stellar mass subsamples,
so that each subsample covers a distinctive portion of the observed stellar mass
distribution. This binning scheme allows us to better constrain the satellite
fraction as a function of stellar mass from the large-scale $w_p$ measurements.

We infer that the LOWZ magnitude and colour cuts select more low-$\ms$ satellite
galaxies from the high-$\mh$ haloes, while the CMASS selection prefers low-$\ms$
satellites from haloes with lower mass. The CMASS preference also increases with
increasing redshift, progressively selecting a higher fraction of low-$\ms$
satellites at higher redshifts. Those inferred behaviours are consistent with
the expectation that the blue fraction of satellites is a increasing function of
redshift, but decreases with halo mass at fixed redshift. In particular, since
the galaxy selection in LOWZ isolates the region of LRGs on the colour-magnitude
diagram, the LOWZ selection prefers satellites from the more massive haloes.
Meanwhile, the CMASS magnitude and colours cuts extend the selection to bluer
populations at higher redshifts, hence the stronger preference of selecting
satellites from low-mass haloes at high redshifts.

Overall, our best-fitting model provides excellent description of the observed
number density and the project clustering of BOSS galaxies, and most important,
a great match to the measured g-g lensing signals on scales below $1\hmpc$ for
all the four redshift samples. On scales larger than $1\hmpc$, similar to the
findings in \citet{Leauthaud2017}, the match is less ideal despite the
relatively large measurement uncertainties, but there is little or no leeway for
adjusting to a better fit without changing the cosmology.  This is because the
large-scale lensing is determined by the galaxy linear bias at any given
cosmology, which is however tightly constrained by the high signal-to-noise
$w_p$ measurements. Therefore, it is unclear whether the large-scale lensing
discrepancy will disappear with improved lensing measurements in the future, or
requires modifications to cosmology or adding even new physics. As to our remedy
to the small-scale lensing discrepancy, we anticipate that the stellar-mass
complete galaxy samples observed by DESI~\citep{DESI2016} and
PFS~\citep{Takada2014} would provide a more conclusive and complete answer to
the validity of our model.

\section*{Acknowledgements}

We thank Johannes Lange for providing the LOWZ lensing measurements, Jesse
Golden-Marx for the assistance with running \texttt{EzGal}, Hong Guo for the
help understanding the BOSS data set, and Ben Wibking for the stimulating
discussions. YZ acknowledges the support by the National Key Basic Research and
Development Program of China (No. 2018YFA0404504), National Science Foundation
of China (11621303, 11873038), the National One-Thousand Youth Talent Program of
China, the STJU start-up fund (No.  WF220407220), and the ``111'' project of the
Ministry of Education under grant No. B20019. YZ also thanks the hospitality of
Cathy Huang during his visit to Zhangjiang Hi-Tech Park that inspired this
project.



\bibliographystyle{mnras}
\bibliography{lil} %



\bsp	
\label{lastpage}

\end{document}